\documentclass[english,twocolumn,floats,showpacs,amsmath,amssymb,prb]{revtex4}
\usepackage[T1]{fontenc}
\usepackage[latin9]{inputenc}
\usepackage{amsmath}
\usepackage{amssymb}
\usepackage{graphicx}

\makeatletter

\providecommand{\tabularnewline}{\\}

\@ifundefined{textcolor}{}
{%
 \definecolor{BLACK}{gray}{0}
 \definecolor{WHITE}{gray}{1}
 \definecolor{RED}{rgb}{1,0,0}
 \definecolor{GREEN}{rgb}{0,1,0}
 \definecolor{BLUE}{rgb}{0,0,1}
 \definecolor{CYAN}{cmyk}{1,0,0,0}
 \definecolor{MAGENTA}{cmyk}{0,1,0,0}
 \definecolor{YELLOW}{cmyk}{0,0,1,0}
 }

\usepackage{epsfig}

\usepackage{babel}

\makeatother

\usepackage{babel}
\begin{document}

\title{The phase diagram of ice Ih, II, and III: a quasi-harmonic study}

\author{R. Ramírez$^{a)}$, N. Neuerburg, and C. P. Herrero\let\oldthefootnote\thefootnote\global\long\def\thefootnote{{a)}}
 \footnotetext{Electronic mail: ramirez@icmm.csic.es}\let\thefootnote\oldthefootnote }

\affiliation{Instituto de Ciencia de Materiales de Madrid (ICMM), Consejo Superior
de Investigaciones Científicas (CSIC), Campus de Cantoblanco, 28049
Madrid, Spain }

\date{{\today}}
\begin{abstract}
The phase diagram of ice Ih, II, and III is studied by a quasi-harmonic
approximation. The results of this approach are compared to phase
diagrams previously derived by thermodynamic integration using path
integral and classical simulations, as well as to experimental data.
The studied models are based on both flexible (q-TIP4P/F) and rigid
(TIP4P/2005, TIP4PQ/2005) descriptions of the water molecule. Many
aspects of the simulated phase diagrams are reasonably reproduced
by the quasi-harmonic approximation. Advantages of this simple approach
are that it is free from the statistical errors inherent to computer
simulations, both classical and quantum limits are easily accessible,
and the error of the approximation is expected to decrease in the
zero temperature limit. We find that the calculated phase diagram
of ice Ih, II, and III depends strongly on the hydrogen disorder of
ice III, at least for cell sizes typically used in phase coexistence
simulations. Either ice II (in the classical limit) or ice III (in
the quantum one) may become unstable depending upon the proton disorder
in ice III. The comparison of quantum and classical limits shows that
the stabilization of ice II is the most important quantum effect in
the phase diagram. The lower vibrational zero-point energy of ice
II, compared to either ice Ih or III, is the microscopic origin of
this stabilization. The necessity of performing an average of the
lattice energy over the proton disorder of ice III is discussed.
\end{abstract}

\pacs{64.60.-i,64.60.De, 63.20.-e, 63.20.Ry}

\maketitle

\section{Introduction\label{sec:intro}}

An outstanding property of water is the diversity of ice phases that
are found in its phase diagram.\cite{dunaeva10} Sixteen different
crystalline ice phases have been identified so far, a number that
is likely to increase in the future. In all phases, except ice X,
the water molecule appears as a well defined entity that is part of
a network of molecules connected by H-bonds. In this network each
water molecule is surrounded by four others in a more or less distorted
tetrahedral coordination. The orientation of each molecule with respect
to its four nearest neighbors satisfies the Bernal-Fowler ice rules.
They state that the H$_{2}$O molecule is oriented so that its two
protons point toward adjacent oxygen atoms and that there must be
one and only one proton between two adjacent oxygen atoms.\cite{bernal33} 

The existence of orientational disorder in the water molecules is
a property of several ice phases. While the oxygen atoms display a
full occupancy ($f$) of their crystallographic positions, the hydrogen
atoms may display a disordered spatial distribution as evidenced by
a fractional occupancy of their lattice sites. Ice Ih, the stable
phase of ice under normal conditions, displays full proton disorder
compatible with the Bernal-Fowler rules, i.e., occupancies of H-sites
of $f=0.5$. However ice II is H-ordered, while ice III is characterized
by a partial proton ordering, i.e., some fractional occupancies of
H-sites are different from 0.5. Order-disorder transitions have been
observed for several pairs of ice phases (Ih-XI, III-IX, V-XIII, VI-XV,
VII-VIII, XII-XIV). The orientational ordering implies a whole reorganization
of the H-bond network that is kinetically unfavorable. In several
disordered phases this transition only occurs after doping with either
bases (in the case of ice Ih) or acids (for ices V, VI, and XII).
The creation of defects provides a mechanism favoring the rearrangement
of the H-bond network.\cite{singer12}

The simulation of the complex phase diagram of water is an interesting
challenge. Large portions of the phase diagram have been simulated
using rigid models (TIP4P,\cite{sanz04} TIP4P/2005,\cite{abascal05}
and TIP4PQ/2005\cite{mcbride12}), and smaller regions using a flexible
water model (q-TIP4P/F)\cite{habershon11}. Let us present a brief
summary of these TIP4P-like models. The TIP4P potential is based on
a point charge description of a rigid water molecule supplemented
by an additional Lennard-Jones interaction between the oxygen centers.
It was parameterized by Jorgensen \textit{et al.} in 1983.\cite{jorgensen83}
An optimized parameterization of the same model was labeled as TIP4P/2005.\cite{abascal05}
Both model potentials have been employed in classical simulations.
Consideration of quantum effects by path integral simulations with
the TIP4P/2005 model led to unphysical results, e.g., ice II was predicted
to be more stable than ice Ih at low temperatures. Then, a small increase
in the point charges was proposed to avoid this problem and the new
parameterization was labeled as TIP4PQ/2005.\cite{mcbride09} An interesting
recent contribution was to add to the rigid TIP4P/2005 model an anharmonic
potential energy term to treat the molecular flexibility of water
in quantum simulations, giving rise to the q-TIP4P/F model.\cite{habershon09}
Summarizing, TIP4P/2005 and TIP4PQ/2005 are rigid water models intended
for its use in classical and quantum simulations, respectively, while
q-TIP4P/F is a flexible model for quantum simulations.

A comprehensive review of the calculation of free energies in water
phases with the thermodynamic integration (TI) method can be found
in Ref. \onlinecite{vega09}. The classical phase diagram of water,
simulated with the TIP4P/2005 model, shows a reasonable qualitative
agreement to the experimental one, in particular in the complex region
of stability of ices Ih, II, III, V, and VI.\cite{abascal05} The
phase diagram of ice Ih, II, and III has been also investigated using
the flexible water model (q-TIP4P/F) in the classical limit.\cite{habershon11}
It was found that ice II is unstable, as its stability region was
occupied by ice III. A plausible explanation of the difference in
the stability of ice II found with the TIP4P/2005 and q-TIP4P/F models
might be that the geometry, dipole and quadrupole moments of the molecules
in the flexible model can vary between the different ice phases, which
is not the case for a rigid model. The simulation of the phase diagram
of water using quantum path integral simulations with the TIP4PQ/2005
model has been recently reported for a temperature range between 140
and 300 K, and pressures up to 1.2 GPa.\cite{mcbride12} The quantum
results were compared to the classical expectation. One of the most
streaking difference between the classical and quantum limits is the
region of stability of ice III, which is much lower (and in better
agreement to the experiment) in the quantum case.

Despite the overall agreement found between simulated and experimental
phase diagrams of water, it is obvious that some properties can not
be reproduced by the employed empirical models. Order-disorder transitions
are a prominent example. It is well documented that these water models
are unable to predict that ice XI (the ferroelectric ordered counterpart
of ice Ih with $Cmc2_{1}$ spatial symmetry) is the most stable ice
phase below 72 K at atmospheric pressure.\cite{buch98} Also the order-disorder
transition between ice VII and VIII is poorly reproduced by the empirical
models.\cite{sanz04} It seems that these effective potentials fail
to describe the energetics of proton rearrangements in ice. Therefore
the location of order-disorder transitions and the identity of the
ordered low temperature phases is inadequately predicted.\cite{singer12}
An interesting question is to what extent the use of \textit{ab initio}
density functional theory (DFT) can improve these limitations. In
this respect, DFT studies of liquid water and ice have revealed serious
differences with experimental data in both diffusive and structural
properties that seems to be related to the subtle contribution of
van der Waals dispersion forces to the cohesive energy of the water
phases. The application of new functionals specially designed to treat
van der Waals interactions is focus of recent interest in modeling
water.\cite{dion04,wang11,santra11}

The quasi-harmonic (QH) approximation (QHA) allows to compute the
partition function of a solid phase as an analytic function of the
crystal volume and the temperature.\cite{srivastava} Some advantages
of this approach are the straightforward derivation of equilibrium
thermodynamic properties, the absence of statistical errors (as opposed
to any simulation method) and the possibility to account for finite
size effects by a Brillouin zone integration of the phonon dispersion
curves, rather than by increasing the size of the cell. The QHA in
combination with \textit{ab initio} DFT has allowed the explanation
of the inverse isotope effect in the crystal volume of ice Ih at atmospheric
pressure.\citep{pamuk12} Also the negative thermal expansion of ice
Ih at low temperatures has been studied by the QHA,\citep{tanaka01}
as well as the elastic moduli and mechanical stability of the H-ordered
ice VIII.\citep{tse99} In addition, the mechanical stability of ice
Ih under pressure has been studied by this approximation.\citep{tse99b}
The ice VII-VIII phase boundary has been studied by a QHA in a 16-molecule
supercell with \textit{ab initio} DFT calculations of total energies
and phonon frequencies.\cite{umemoto10} 

The validity of the QHA is restricted by the possible presence of
anharmonic effects beyond those included in the approximation. Thus,
a direct check of the QHA is the comparison to numerical simulations
that fully consider the anharmonicity of the interatomic interactions.
The QHA prediction of the volume, enthalpy, kinetic energy, and heat
capacity, of ice Ih, II, and III has been compared to both quantum
and classical simulations using the q-TIP4P/F model.\cite{ramirez12}
The comparison in a ($T,P$) range up to 300 K and 1 GPa shows a remarkable
overall agreement for the three ice phases. An interesting aspect
of the QHA is that it is sensible enough to predict differences in
the anharmonicity of different water models, that are in agreement
to the corresponding fully anharmonic simulations. For example, the
QHA predicts that the thermal expansion of ice Ih at low temperatures
is negative for the q-TIP4P/F and TIP4P models but positive (or slightly
negative) for the TIP5P and ST2 potentials.\cite{ramirez12,koyama04}
Moreover, the isotope effect in the crystal volume of ice Ih is predicted
by the QHA to be anomalous (as in the experiment) with a DFT functional,
but normal with the q-TIP4P/F model.\cite{pamuk12} We stress that
these QHA predictions of \textit{anharmonic effects} are in agreement
to results of available computer simulations. It may be somewhat surprising
that the \textit{simple} QHA approximation is able to reproduce the
anharmonicity of complex ice phases with a similar accuracy as that
shown for solids with much simpler crystal structures such as noble
gases and elemental semiconductors (Si, Ge).\cite{noya97,herrero00,herrero05} 

The purpose of the present paper is to check the capability of the
QHA to predict the phase diagram of ice Ih, II, and III. The layout
of the manuscript is as follows. A summary of the employed computational
conditions is presented in Sec. \ref{sec:Computational-conditions}.
The generation of the ice structures is introduced in Sec. \ref{sec:ice structures}.
The QH phase diagram is studied for the flexible q-TIP4P/F model in
Sec. \ref{sec:flexible-q-TIP4P/F-model}. The results for the rigid
models, TIP4P/2005 and TIP4PQ/2005, are presented in Secs. \ref{sec:Rigid-TIP4P/2005-model}
and \ref{sec:Rigid-TIP4PQ/2005-model}, respectively. Our main focus
of interest is the influence of proton disorder in the calculated
phase diagram, the comparison of the QHA to previous simulation results,
and the differences between the quantum and classical limits. The
necessity of performing disorder averaging is discussed in Sec. \ref{sec:Average-over-proton}.
The paper closes with the conclusions.

\section{Computational conditions\label{sec:Computational-conditions}}

The QHA employed for the ice phases has been introduced in Ref. \onlinecite{ramirez12}.
We present here a brief summary. The Helmholtz free energy of an ice
phase with $N$ water molecules in a cell of volume $V$ and at temperature
$T$ is given by
\begin{equation}
F(V,T)=U_{S}(V)+F_{v}(V,T)-TS_{H}\;,\label{eq:f_v_t}
\end{equation}
 where $U_{S}(V)$ is the static zero-temperature classical energy,
i.e., the minimum of the potential energy when the volume of the ice
cell is $V$. The entropy $S_{H}$ is related to the disorder of hydrogen
and it vanishes for ordered ice phases as ice II. $S_{H}$ was estimated
by Pauling for fully disordered phases as \citep{pauling35}
\begin{equation}
S_{H}=Nk_{B}\ln\frac{3}{2}\;.\label{eq:entropy}
\end{equation}
 $F_{v}(V,T)$ is the vibrational contribution to $F$. In the quantum
limit is given by
\begin{equation}
F_{v}(V,T)=\sum_{k}\left(\frac{\hbar\omega_{k}}{2}+\frac{1}{\beta}\ln\left[1-\exp\left(-\beta\hbar\omega_{k}\right)\right]\right)\;.\label{eq:fv_q}
\end{equation}
Here $\beta$ is the inverse temperature: $1/k_{B}T$. $\omega_{k}$
are the wavenumbers of the harmonic lattice vibrations for the volume
$V$, with $k$ combining the phonon branch index and the wave vector
within the Brillouin zone. In the classical limit the vibrational
contribution amounts to
\begin{equation}
F_{v,cla}(V,T)=\sum_{k}\frac{1}{\beta}\ln\left(\beta\hbar\omega_{k}\right)\;.
\end{equation}
 The Gibbs free energy, $G(T,P),$ is obtained by seeking for the
volume, $V_{min}$, that minimizes the function $F(V,T)+PV$, as
\begin{equation}
G(T,P)=F(V_{min},T)+PV_{min}\;.
\end{equation}
The implementation of the QHA for an ice phase follows these steps:\cite{ramirez12} 

$i)$ Find the reference cell that minimizes the static energy $U_{S}$.
This minimization implies optimization of both cell shape and atomic
positions. The resulting volume is $V_{ref}$ and the corresponding
static energy $U_{S,ref}$.

$ii)$ Select a grid of 50 volumes in a range of interest $\left[V_{min},V_{max}\right]$.
The ice cell with volume $V_{i}$ is set by isotropic scaling of the
reference cell. Subsequently, each ice cell is held fixed while minimizing
the static energy $U_{S}(V_{i})$ with respect to the atomic positions.
The crystal phonons, $\omega_{k}(V_{i})$, are obtained after the
minimization. 

$iii)$ Calculate the function $F(V_{i},T)$ by Eq. (\ref{eq:f_v_t}).
The minimum of $F(V_{i},T)$ as a function of $V$ is determined by
a fit to a 5th degree polynomial in $V$. 

The phase diagram of the ice phases is derived by a brute force method,
i.e., given a state point $(T,P)$ one calculates the Gibbs free energy
of all ice phases and then the stable phase is selected as the one
with the lowest value of $G$. 

The crystal phonon calculation has been performed by the small-displacement
method.\citep{kresse95,alfe01} For the flexible water model the atomic
displacement employed in this work is $\delta x=10^{-6}$ $\textrm{\AA}$
along each Cartesian direction. For the rigid water models the molecular
displacements imply translations (by $10^{-6}$ $\textrm{\AA}$ along
the Cartesian directions) and rotations (by $10^{-7}$ rad along the
Cartesian axes) of the rigid molecules. See Ref. {[}\onlinecite{venkataraman70}{]}
for a full account of the calculation of the external phonon modes
associated to rigid units. We have used a $\Gamma$ sampling (\textbf{$\mathbf{k}=\mathbf{0}$})
of the crystal phonons, as this condition is implicitly assumed in
simulation studies using periodic boundary conditions.\cite{ramirez12}
The Lennard-Jones interaction between oxygen centers was truncated
at a distance of  $r_{c}=8.5\textrm{ \AA}$, and standard long-range
corrections for both potential energy and pressure were computed assuming
that the pair-correlation function is unity for $r>r_{c}$.\cite{johnson93}
Long-range electrostatic potential and forces were calculated with
the Ewald method. 

As a check for the assumption of isotropic expansion of the reference
cell made in the step \textit{ii}), we have performed a QHA of the
Gibbs free energy of ice II by relaxing this constraint. To this aim
we have derived a set of 50 cell volumes $V_{i}$ by performing an
optimization of both cell shapes and atomic positions at 50 different
pressures in the range {[}-1.7, 3{]} GPa. The new result for the free
energy of ice II reveals that the assumption of isotropic scaling
of the reference cell modifies the values of $G$ by less than 0.01
kJ/mol. This change in $G$ has a small effect in the phase diagram
of ice reflected by rigid shifts of the calculated coexistence lines
of ice II by less than 2~K.

\section{Ice structures\label{sec:ice structures}}

Supercells of similar size to those employed in recent simulations\cite{habershon11,mcbride12}
have been used in the QH derivation of the phase diagram. The number
of molecules were $N$= 288 for ice Ih, and $N$= 324 for ice II and
III. The ice Ih cell was orthorhombic with parameters $(4\mathbf{a}_{1},3\sqrt{3}\mathbf{a}_{1},3\mathbf{a}_{3})$,
with $(\mathbf{a}_{1},\mathbf{a}_{3})$ being the standard hexagonal
lattice vectors of ice Ih.\citep{hayward87} Ice II and III were studied
by $3\times3\times3$ supercells of the crystallographic cell, which
belong to the rhombohedral and the tetragonal crystal systems, respectively.\citep{kamb71,lobban00}
While ice II is proton ordered both ice Ih and III display orientational
disorder of the water molecules. The algorithm proposed by Buch \textit{et
al.} was applied for the random generation of full proton disordered
structures ($f=0.5$) with vanishing cell dipole moment.\cite{buch98}
In the case of ice III, the neutron diffraction experiments show that
only 1/3 of the H-sites has $f=0.5$, while the other 2/3 display
occupancies of $f=0.35$ and $f=0.65$, respectively. \cite{lobban00}
The Buch's algorithm has been modified for the generation of random
structures with partial H-disorder, i.e., having fractional occupancies
different from $f=0.5.$\cite{macdowell04}

An interesting practical question is the importance of proton disorder
in the evaluation of the partition function of the H-disordered phases,
using a single H-isomer.
This point has not been addressed earlier in the simulation of phase
diagrams of ice using either TIP4P, TIP4P/2005, TIP4PQ/2005, or q-TIP4P/F
models.\cite{sanz04,abascal05,habershon11,mcbride12} These simulations
used a \textit{single H-isomer} of the disordered phases (ice Ih and
III), generated by a \textit{random} algorithm. Moreover, the H-isomer
for ice III was selected either with \textit{partial H-disorder} in
the earlier simulation with the rigid TIP4P model\cite{sanz04} or
with \textit{full H-disorder} in more recent simulations with the
q-TIP4P/F and TIP4PQ/2005 potentials.\cite{habershon11,mcbride09}
The question to be addressed here is how large might be the influence
of the selected H-isomer in the calculated phase diagram. 

\begin{figure}
\vspace{-1.8cm}
\includegraphics[width= 9cm]{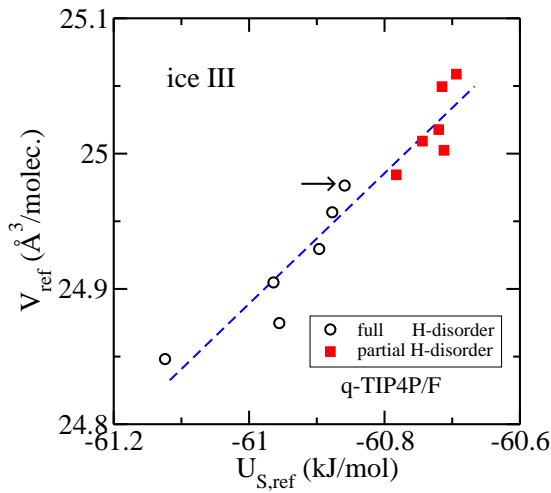}
\vspace{-0.9cm}
\caption{Minimum static energy and corresponding cell volume of randomly
generated
H-isomers of ice III. The H-isomers display either full (open circles)
or partial (closed squares) H-disorder. The results correspond to
the q-TIP4P/F model for a 324-molecule supercell. An arrow points
to the H-isomer with full H-disorder and largest energy. This particular
isomer is employed in the calculation of phase diagrams with different
water models. The line is a linear fit to the data.}
\label{fig:u0_v}
\end{figure}

To this aim we have generated a random set of six H-isomers with full
H-disorder and vanishing cell dipole moment for ice III. The result
of their energy minimization with the flexible q-TIP4P/F potential
is represented in Fig. \ref{fig:u0_v}. The volume, $V_{ref}$, and
the corresponding minimized potential energy, $U_{S,ref}$, is displayed
by open circles for each H-isomer. We note that the static energy,
$U_{S,ref}$, of the six H-isomers spreads in an energy window of
about 0.3 kJ/mol. The volume, $V_{ref}$, and the minimized potential
energy, $U_{S,ref}$, are related in a nearly linear way. A second
set of six random isomers with vanishing cell dipole moment has been
generated by imposing the partial H-disorder encountered in the diffraction
experiment of ice III.\cite{lobban00} The results of the energy minimization
for the partially disordered structures are presented as closed squares
in Fig. \ref{fig:u0_v}. All isomers having partial H-disorder display
\textit{larger} static energy than the isomers with full H-disorder. 

Similar behavior to that shown in Fig. \ref{fig:u0_v} is found if
the minimization of the energy of the H-isomers is performed with
the rigid TIP4P/2005 model. The main difference is that the dispersion
of the $U_{S,ref}$ values increases slightly. Thus, two main conclusions
can be derived from the results of the energy minimization in Fig.
\ref{fig:u0_v}. The first is that the energetics associated to full
versus partial H-disorder in ice III is incorrectly described by effective
TIP4P-like models, i.e., full disorder is predicted to be more stable
than partial one. Note that the configurational entropy $S_{H}$ will
help to stabilize further the full disordered ice at any finite temperature,
since $S_{H}$ in this case is larger than for the partial H-disorder.
This behavior is in contradiction to the partial H-disorder experimentally
found for ice III.\cite{lobban00} Our result is in line with the
reported limitations of these effective potentials to reproduce the
energetics of the H-bond rearrangement in the order-disorder transition
of ice Ih-XI and VII-VIII.\cite{buch98,singer12} 

Our second conclusion is that the large dispersion in $U_{S,ref}$
obtained for ice III using cells with 324 molecules must affect the
phase diagram whenever it is calculated with a \textit{single random}
H-isomer. The energy dispersion is caused by the proton disorder in
the H-isomers. The sampling of this (large) dispersion of static cell
energies with a single H-isomer can be considered the origin of a
\textit{finite size effect}. In the thermodynamic limit, the energy
distribution of $U_{S,ref}$ should approximate a delta function centered
at the energy corresponding to the most probable H-bond distribution.
Therefore, the finite size effect caused by the insufficient sampling
of the proton disorder (or the $U_{S,ref}$ energies) with a single
H-isomer is expected to decrease as the size of the cell increases.
An alternative way to reduce this finite size effect is to make a
disorder averaging of the lattice energy of the employed ice cell.
This point will be commented in Sec. \ref{sec:Average-over-proton}.

In the case of ice Ih we have also generated a random set of six H-disordered
structures. However, in this case the static energy, $U_{S,ref}$,
of the H-isomers varies in a rather small energy window (of about
0.01 kJ/mol for the q-TIP4P/F model) and the corresponding volume
changes by less than 0.04\%. Thus, finite size effects related to
the H-disorder are expected to be much lower in ice Ih than in ice
III.

\begin{table*}
\caption{Volume ($V_{ref}$) and static energy ($U_{S,ref}$) of the minimum
energy configuration of the studied ice phases. The quantum QH results
for the volume ($V_{0}$), static energy ($U_{S,0}$), zero-point
energy ($U_{Z,0}$), and internal energy ($U_{0}$) are also given
at $T=0$ and $P=0$. The data for ice III correspond to the H-isomer
labeled by an arrow in Fig. \ref{fig:u0_v}. All results were derived
with the q-TIP4P/F model. The last two columns show the difference
with the data of ice II. $\left[V_{min},V_{max}\right]$ is the volume
interval studied by the QHA for each phase.}
\vspace{4mm}
\begin{tabular}{cr@{\extracolsep{0pt}.}lr@{\extracolsep{0pt}.}
                 lr@{\extracolsep{0pt}.}lrr}
\hline
$X$ (q-TIP4P/F) & \multicolumn{2}{c}{$\quad$Ih} & \multicolumn{2}{c}{II}
& \multicolumn{2}{c}{III} & $\Delta X$ (Ih-II) & $\Delta X$
(III-II)\tabularnewline
\hline
$V_{ref}$ ($\textrm{\AA}^{3}/$molec.) & $\quad$30&96 & 24&14 & 24&99 &
6.82 & 0.85\tabularnewline
$U_{S,ref}$ (kJ/mol) & $\quad$-61&98 & -60&84 & -60&86 & -1.14 &
-0.02\tabularnewline
$V_{0}$ ($\textrm{\AA}^{3}/$molec.) & $\quad$32&23 & 25&11 & 25&90 &
7.12 & 0.79\tabularnewline
$U_{S,0}$ (kJ/mol) & $\quad$-61&74 & -60&60 & -60&77 & -1.14 &
-0.17\tabularnewline
$U_{Z,0}$ (kJ/mol) & $\quad$68&75 & 68&08 & 68&72 & 0.67 &
0.64\tabularnewline
$U_{0}$ (kJ/mol) & $\quad$7&01 & 7&47 & 7&95 & -0.46 &
0.47\tabularnewline
$V_{min}$ ($\textrm{\AA}^{3}/$molec.) & $\quad$29&47 & 21&75 & 22&48 &
& \tabularnewline
$V_{max}$ ($\textrm{\AA}^{3}/$molec.) & $\quad$35&05 & 27&31 & 28&22 &
& \tabularnewline
\hline
\end{tabular}
\label{tab:ices_ref_fle}
\end{table*}

A comparison of $V_{ref}$ and $U_{S,ref}$ calculated with the q-TIP4P/F
model for ice Ih, II, and III are given in Tab. \ref{tab:ices_ref_fle}.
The data for ice III correspond to the H-isomer labeled with an arrow
in Fig. \ref{fig:u0_v}. The classical internal energy at zero temperature
and pressure ($T=0$, $P=0$) is $U_{S,ref}$. The QH result in the
quantum limit for the ice volume ($V_{0}$), static energy ($U_{S,0}$),
and zero-point energy ($U_{Z,0}$) at $T=0$ and $P=0$ are also summarized
in Tab. \ref{tab:ices_ref_fle}. The zero-point energy, $U_{Z,0}$,
is calculated as 
\begin{equation}
U_{Z,0}=\sum_{k}\frac{\hbar\omega_{k}(V_{0})}{2}\;.\label{eq:Uz}
\end{equation}
Note that the zero-point energy of ice II is lower than that of ice
Ih and III. In the quantum limit, the internal energy of the ice phases
at $T=0$ and $P=0$ is 
\begin{equation}
U_{0}=U_{S,0}+U_{Z,0}\;.
\end{equation}
The ice structures studied in Tab. \ref{tab:ices_ref_fle} have been
analyzed also with the rigid TIP4P/2005 and TIP4PQ/2005 models. The
corresponding results are presented in Tabs. \ref{tab:ice_ref_2005}
and \ref{tab:ice_ref_Q2005}. Note that the zero-point energy ($U_{Z,0}$)
of the rigid models is about four times smaller than that of flexible
water because of the neglect of intramolecular motion.

\begin{table*}
\caption{Volume and energies of the studied ice phases as derived with
the
rigid TIP4P/2005 model at $T=0$ and $P=0$. The ice structures and
variable labels are the same as those used in Tab.
\ref{tab:ices_ref_fle}. }
\vspace{4mm}
\begin{tabular}{cr@{\extracolsep{0pt}.}lr@{\extracolsep{0pt}.}lr@{\extracolsep{0pt}.}lrr}
\hline
$X$ (TIP4P/2005) & \multicolumn{2}{c}{$\quad$Ih} &
\multicolumn{2}{c}{II} & \multicolumn{2}{c}{III} & $\Delta X$ (Ih-II) &
$\Delta X$ (III-II)\tabularnewline
\hline
$V_{ref}$ ($\textrm{\AA}^{3}/$molec.) & $\quad$31&34 & 24&30 & 25&26 &
7.04 & 0.96\tabularnewline
$U_{S,ref}$ (kJ/mol) & $\quad$-62&99 & -62&13 & -61&86 & -0.86 &
0.27\tabularnewline
$V_{0}$ ($\textrm{\AA}^{3}/$molec.) & $\quad$33&20 & 25&71 & 26&76 &
7.59 & 1.05\tabularnewline
$U_{S,0}$ (kJ/mol) & $\quad$-62&45 & -61&66 & -61&63 & -0.79 &
0.03\tabularnewline
$U_{Z,0}$ (kJ/mol) & $\quad$16&15 & 15&10 & 16&29 & 1.05 &
1.19\tabularnewline
$U_{0}$ (kJ/mol) & $\quad$-46&30 & -46&56 & -45&33 & 0.26 &
1.23\tabularnewline
\hline
\end{tabular}
\label{tab:ice_ref_2005}
\end{table*}

\begin{table*}
\caption{Volume and energies of the studied ice phases as derived with
the
rigid TIP4PQ/2005 model at $T=0$ and $P=0$. The ice structures and
variable labels are the same as those used in Tab.
\ref{tab:ices_ref_fle}. }
\vspace{4mm}
\begin{tabular}{cr@{\extracolsep{0pt}.}lr@{\extracolsep{0pt}.}lr@{\extracolsep{0pt}.}lrr}
\hline
$X$ (TIP4PQ/2005) & \multicolumn{2}{c}{$\quad$Ih} &
\multicolumn{2}{c}{II} & \multicolumn{2}{c}{III} & $\Delta X$ (Ih-II) &
$\Delta X$ (III-II)\tabularnewline
\hline
$V_{ref}$ ($\textrm{\AA}^{3}/$molec.) & $\quad$30&67 & 23&85 & 24&92 &
6.82 & 1.07\tabularnewline
$U_{S,ref}$ (kJ/mol) & $\quad$-68&90 & -67&57 & -67&47 & -1.33 &
0.10\tabularnewline
$V_{0}$ ($\textrm{\AA}^{3}/$molec.) & $\quad$32&51 & 25&16 & 26&32 &
7.35 & 1.16\tabularnewline
$U_{S,0}$ (kJ/mol) & $\quad$-68&35 & -67&11 & -67&24 & -1.24 &
-0.13\tabularnewline
$U_{Z,0}$ (kJ/mol) & $\quad$17&06 & 15&91 & 17&19 & 1.15 &
1.28\tabularnewline
$U_{0}$ (kJ/mol) & $\quad$-51&30 & -51&20 & -50&05 & -0.10 &
1.15\tabularnewline
\hline
\end{tabular}
\label{tab:ice_ref_Q2005}
\end{table*}

\section{Flexible q-TIP4P/F model\label{sec:flexible-q-TIP4P/F-model}}

In this section the QH phase diagram of ice Ih, II, and III is derived
with the q-TIP4P/F model in both classical and quantum limits. Studied
temperatures are in the interval {[}0,$300\;\mathrm{K}${]} and pressures
in the range {[}0, 0.35 GPa{]}. The calculation is done for each of
the six random H-isomers of ice III having full H-disorder. A comparison
to available TI simulations is provided in the classical 
limit.\cite{habershon11}

\subsection{Classical limit}

\begin{figure}
\vspace{-0.5cm}
\includegraphics[width= 9cm]{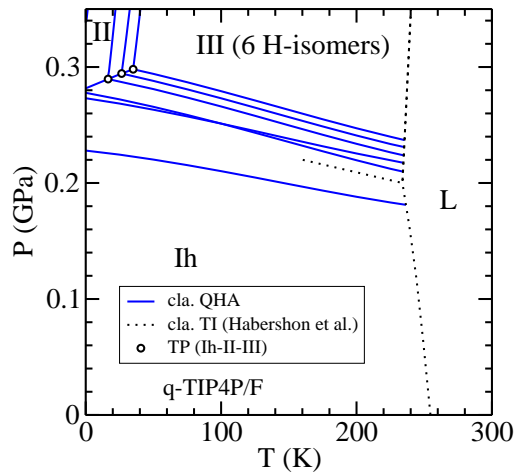}
\vspace{-0.9cm}
\caption{Phase
diagram of ice Ih-II-III calculated with the q-TIP4P/F model
in the classical limit. The full lines show the QH results. The QH
calculation is done for six randomly chosen H-isomers having full
H-disorder in a cell with 324 water molecules. The dotted lines are
the classical TI results of Ref. \onlinecite{habershon11} that include
also the boundary with the liquid (L) phase. Circles show the position
of the triple point (TP) for ice Ih-II-III.}
\label{fig:classical}
\end{figure}

The QH phase diagram of ice Ih, II, and III calculated in the classical
limit is plotted in Fig. \ref{fig:classical}. Coexistence lines are
displayed for each of the six studied H-isomers of ice III as continuous
curves. We find that the finite size effect related to the H-disorder
in ice III is important. In particular, the area where ice III is
stable strongly depends upon the H-isomer. The dotted lines in Fig.
\ref{fig:classical} show the coexistence lines for ice Ih-III, Ih-liquid,
and III-liquid as derived from the classical TI simulations of Ref.
\onlinecite{habershon11} with the q-TIP4P/F model. The coexistence
line Ih-III is parallel to our QHA results. 

\begin{figure}
\vspace{-0.6cm}
\includegraphics[width= 9cm]{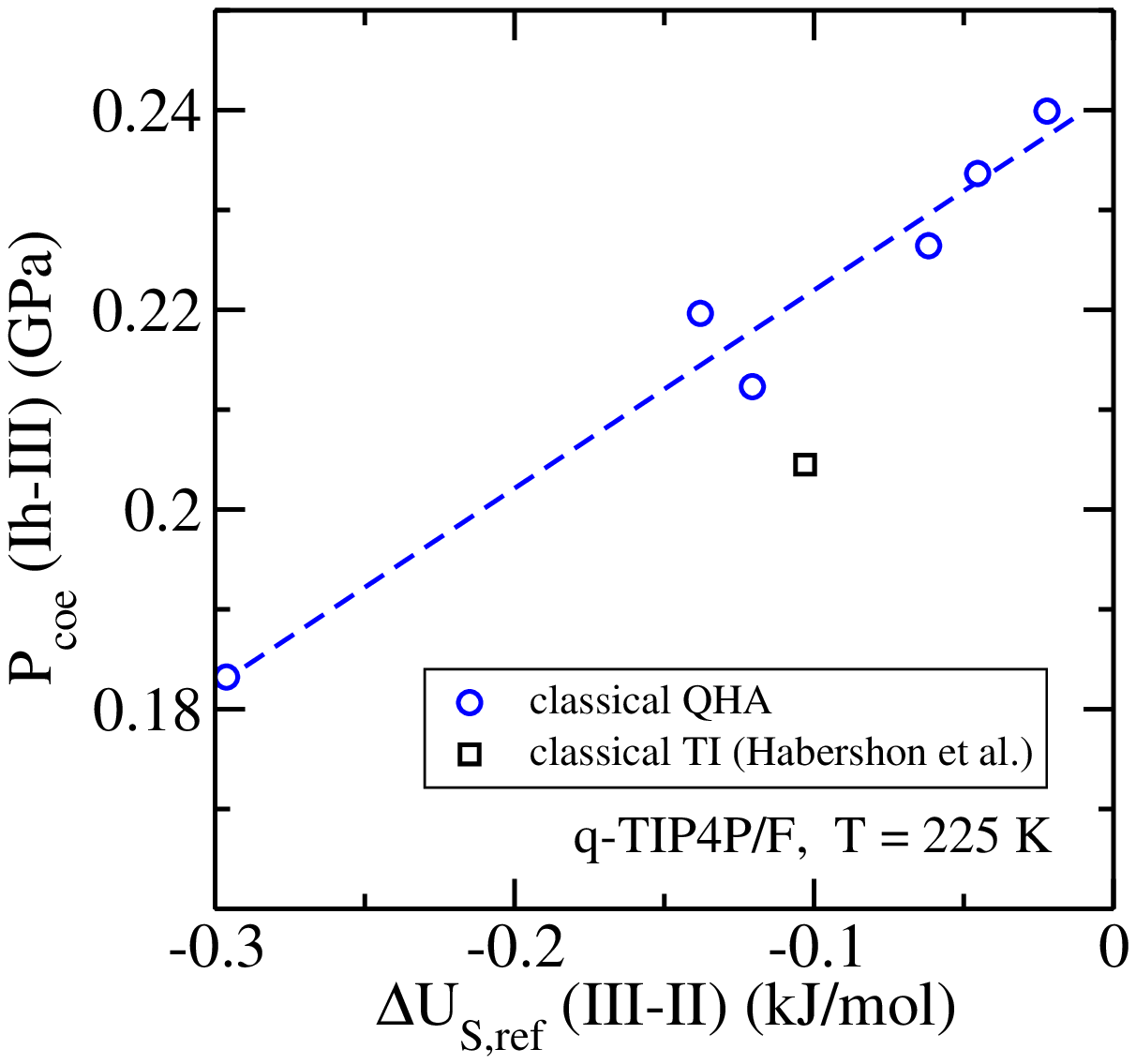}
\vspace{-0.9cm}
\caption{Coexistence pressure for ice Ih and III at 225~K as a function
of the relative static energy of the H-isomers of ice III. The static
energy of ice II has been taken as zero of the energy scale. Circles
are results derived by the QHA using the q-TIP4P/F model. The square
is the result of Ref. \onlinecite{habershon11} based on the TI method
with the same water potential. The line is a linear fit to the QH
data.}
\label{fig:pc1-3_u}
\end{figure}

For the various H-isomers, the ice Ih-III phase boundary is shifted
by a nearly constant pressure. The dispersion of the static energy,
$U_{S,ref}$, is the factor responsible for the different phase behavior
of the H-isomers. Vibrational contributions to the free energy are
however similar. The coexistence pressure for the Ih-III transition
at the temperature of $T=225$~K is represented in Fig. \ref{fig:pc1-3_u}
as a function of the relative static energy, $\Delta U_{S,ref}$,
of the H-isomers of ice III. The relative energy is calculated with
respect to the minimum potential energy of ice II, to allow for a
comparison to available literature data.\cite{habershon11} The coexistence
pressure varies in the interval 0.18-0.24 GPa at 225~K. There appears
an approximate linear relation between the coexistence pressure and
the static energy of the ice III isomer. The coexistence pressure
reported using TI simulations deviates by less than 0.02 GPa (about
8\%) from the linear fit based upon the QH results. This small deviation
suggests that the QHA is reasonably realistic even at this relatively
high temperature ($T=225$~K). The difference between the QHA and
the TI simulations is caused by the presence of anharmonic effects
not included in the QHA and also by the use of different H-isomers
in both calculations. Unfortunately it is not possible to quantify
the separate influence of both factors.

An interesting result from our QH phase diagram is that, for cells
with 324 molecules, ice II may be either stable at low $T$ or unstable
at all temperatures in the classical limit of the q-TIP4P/F water
model. We find in Fig. \ref{fig:classical} that ice II is unstable
in the whole ($T,P$) region if the phase diagram is calculated with
any of the three most stable H-isomers of ice III. Thus, the stability
of ice II is determined by the static energy, $U_{S,ref}$, of the
H-isomer of ice III. 

Some differences in the phase diagrams calculated with TIP4P-like
models might be caused by the differences in the static energy, $U_{S,ref}$,
of the single H-isomer chosen to represent ice III. For example, the
classical phase diagram for the rigid TIP4P model was calculated with
a H-isomer of ice III with partial H-disorder.\cite{sanz04} We have
seen in Fig. \ref{fig:u0_v} that partial H-disorder is less stable
(it has higher energy) than full H-disorder for TIP4P-like models.
Thus, this choice helps to increase the stability region of ice II.
On the contrary, the phase diagram for the flexible q-TIP4P/F model
was calculated with full H-disorder for ice III. Here the increased
stabilization of ice III (see Fig. 1) plays an important role for
the reported instability of ice II.\cite{habershon11} Our QH calculation
strongly suggests that the ice II instability is not a deficiency
of the q-TIP4P/F model, but a \textit{finite size effect} related
to the particular H-isomer randomly selected for the simulation.

We turn now to the calculation of the QH phase diagram of ice Ih,
II, and III for the quantum limit of the q-TIP4P/F model. In this
case there are quantum TI results for the melting of ice Ih at atmospheric
pressure,\cite{ramirez10,habershon11b} but not for the coexistence
between different ice phases.

\subsection{Quantum limit}

\begin{figure}
\vspace{-1.8cm}
\includegraphics[width= 9cm]{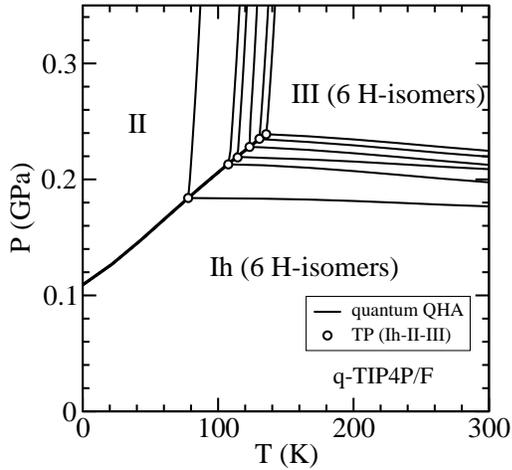}
\vspace{-0.9cm}
\caption{QH phase diagram of ice Ih-II-III calculated with the q-TIP4P/F
model in the  quantum limit. The multiple coexistence lines II-III and Ih-III
show the results for the six studied H-isomers of ice III. The phase
boundary Ih-II is a plot of six superimposed lines, each one calculated
with a different random H-isomer of ice Ih. Circles show the position
of the triple point (TP) for the ice phases.}
\label{fig:T}
\end{figure}

The QH phase diagram of ice Ih, II, and III in the quantum limit is
plotted in Fig. \ref{fig:T}. Finite size effects related to proton
disorder are very important for ice III ($N=324$). As in the classical
limit, this effect is related to the differences in the static energy,
$U_{S,ref}$, of the H-isomers of ice III. Vibrational contributions
to the free energy are however similar, and then the coexistence lines
calculated for the ice III isomers are nearly parallel.

In contrast, the QHA reveals that finite size effects related to H-disorder
are vanishingly small for ice Ih ($N=288$). The coexistence line
Ih-II was calculated for each of the six randomly generated H-isomers
of ice Ih. In this case, the six calculated Ih-II phase boundaries
appear superposed as a unique line at the scale of the figure. The
coexistence lines Ih-III have been displayed for a single H-isomer
of ice Ih.

Ice II is always a stable phase in the quantum phase diagram independently
of the employed H-isomer of ice III. A triple point Ih-II-III appears
for all studied H-isomers, in contrast to the classical results in
Fig. \ref{fig:classical}. The triple point temperature, $T_{TP}$,
is found in an interval 75-136 K depending upon the H-isomer of ice
III. The triple point pressure, $P_{TP}$, appears in the interval
0.18-0.24 GPa. The magnitude of these intervals provides a quantitative
estimation of the influence of the finite size effect of the proton
disorder of ice III in the calculated phase diagram. We find that
both quantities ($T_{TP},P_{TP}$) show a linear dependence as a function
of the static energy, $U_{S,ref}$, of ice III, in a way very similar
to that shown in Fig. \ref{fig:pc1-3_u} for the coexistence pressure
between ice Ih and III.

\subsection{Comparison of quantum and classical limits}

\begin{figure}
\vspace{-0.8cm}
\includegraphics[width= 9cm]{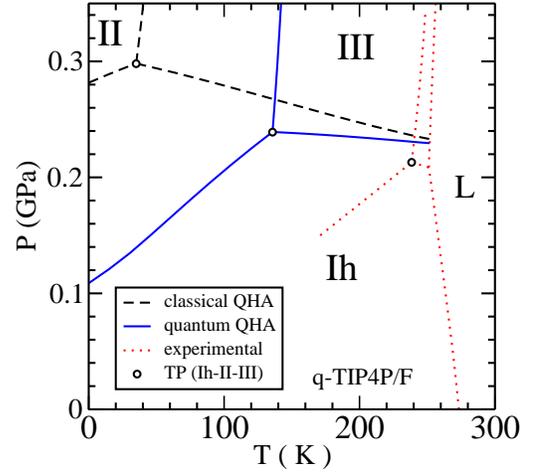}
\vspace{-0.8cm}
\caption{Comparison of the QH phase diagram of ice Ih, II, and III in
the classical
and quantum limits. Dotted lines correspond to experimental data from
Ref. \onlinecite{dunaeva10} that include the liquid (L) phase. Results
derived with the q-TIP4P/F model. The calculation was performed with
the H-isomer of ice III labeled with an arrow in Fig. \ref{fig:u0_v}.
Circles show the position of the triple point for ice Ih-II-III.}
\label{fig:Ae}
\end{figure}

The quantum and classical limits of the QH phase diagram for the q-TIP4P/F
model are compared in Fig. \ref{fig:Ae}. The H-isomer of ice III
indicated by an arrow in Fig. \ref{fig:u0_v} has been arbitrarily
chosen for this comparison. The static energy of this H-isomer is
the closest one to that of partial H-disorder structures. The most
conspicuous quantum effect is the increased stability of ice II. The
triple point Ih-II-III is found classically at (35 K, 0.3 GPa), while
the quantum limit is (136 K, 0.24 GPa), i.e. a shift of about 100
K and -0.06 GPa due to the consideration of quantum vibrational effects.
The experimental boundaries in this region of the phase diagram are
shown in Fig. \ref{fig:Ae} by dotted lines. The experimental triple
point Ih-II-III is found at (239 K, 0.21 GPa).\cite{dunaeva10}

\subsubsection{Coexistence Ih-II\label{sub:Coexistence-Ih-II}}

\begin{figure}
\vspace{-1.5cm}
\includegraphics[width= 9cm]{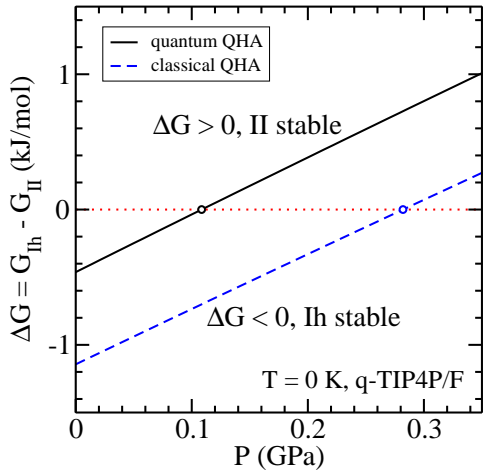}
\vspace{-0.9cm}
\caption{Gibbs free energy difference of ice Ih and II as a function of
$P$ at $T=0$ K. Coexistence conditions are labeled by open symbols 
($\Delta G=0$).
Results derived with the q-TIP4P/F model in both classical and quantum
limits.}
\label{fig:gI_II}
\end{figure}

In the quantum limit, ice II occupies a large portion of the region
of stability found classically for ice Ih. Therefore, the coexistence
line Ih-II is shifted to lower pressures with respect to the classical
one (see Fig. \ref{fig:Ae}). The Gibbs free energy difference between
ice Ih and II, 
\begin{equation}
\Delta G=G_{Ih}-G_{II}\;,\label{eq:d}
\end{equation}
is plotted in Fig. \ref{fig:gI_II} as a function of the pressure
at $T=0$ K. The zero temperature condition implies that
\begin{equation}
\Delta G=\Delta U+P\Delta V\;.\label{eq:u+pv}
\end{equation}
The plot of $\Delta G$ in Fig. \ref{fig:gI_II} is nearly linear
in $P$. This fact, in the light of Eq. (\ref{eq:u+pv}), implies
that both $\Delta U$ and $\Delta V$ vary slowly with $P$ in the
studied pressure interval. 

The main difference between the quantum and classical result for $\Delta G$
is the value of the ordinate at the origin, $\Delta U_{0}$. Fig.
\ref{fig:gI_II} shows that in the classical limit 
\begin{equation}
\Delta U_{0,cla}=-1.14\;\mathrm{kJ/mol}\left(\equiv\Delta U_{S,ref}\right)\;,
\end{equation}
a value that corresponds to the difference in the static energies,
$U_{S,ref},$ of ice Ih and II (see Tab. \ref{tab:ices_ref_fle}).
In the quantum limit the ordinate at the origin is
\begin{equation}
\Delta U_{0}=-0.46\;\mathrm{kJ/mol}\left(\equiv\Delta U_{S,0}+\Delta U_{Z,0}\right)\;.
\end{equation}
The quantum result differs from $\Delta U_{0,cla}$ by an energy increment
that essentially corresponds to the difference in the zero-point energy,
$U_{Z,0},$ of ice Ih and II at $T=0$ and $P=0$. The data in Tab.
\ref{tab:ices_ref_fle} show that
\begin{equation}
\Delta U_{Z,0}=U_{Z,0,Ih}-U_{Z,0,II}=0.67\;\mathrm{kJ/mol}\;,\label{eq:d_uz}
\end{equation}
i.e., the zero-point energy of ice II is 0.67 kJ/mol lower than that
of ice Ih. This is the physical reason for the quantum shift in the
coexistence pressure of the Ih-II transition (abscissa of the open
circles in Fig. \ref{fig:gI_II}) and the origin of the increased
stabilization of ice II in the quantum phase diagram. 

\begin{table}
\caption{Average of the wavenumbers obtained with the q-TIP4P/F model
for the
volume $V_{0}$ of the studied ices. $V_{0}$ is the equilibrium volume
in the quantum limit at $T=0$ and $P=0$. The last two columns show
the ratio of the wavenumbers with respect to the data of ice II. }
\vspace{4mm}
\begin{tabular}{cr@{\extracolsep{0pt}.}lr@{\extracolsep{0pt}.}lr@{\extracolsep{0pt}.}lr@{\extracolsep{0pt}.}lr@{\extracolsep{0pt}.}l}
\hline
$\overline{\omega}_{k}$ (cm$^{-1})$ & \multicolumn{2}{c}{$\quad$Ih} &
\multicolumn{2}{c}{II} & \multicolumn{2}{c}{III} &
\multicolumn{2}{c}{$\quad$Ih/II} &
\multicolumn{2}{c}{$\quad$III/II}\tabularnewline
\hline
$3N$ translations & \multicolumn{2}{c}{$\quad$186} &
\multicolumn{2}{c}{176} & \multicolumn{2}{c}{189} & $\quad$1&06 &
$\quad$1&07\tabularnewline
$3N$ librations & \multicolumn{2}{c}{$\quad$747} &
\multicolumn{2}{c}{684} & \multicolumn{2}{c}{738} & $\quad$1&09 &
$\quad$1&08\tabularnewline
$N$ bending & \multicolumn{2}{c}{$\quad$1684} & \multicolumn{2}{c}{1673}
& \multicolumn{2}{c}{1682} & $\quad$1&01 & $\quad$1&01\tabularnewline
$2N$ stretching & \multicolumn{2}{c}{$\quad$3506} &
\multicolumn{2}{c}{3565} & \multicolumn{2}{c}{3512} & $\quad$0&98 &
$\quad$0&99\tabularnewline
\hline
\end{tabular}
\label{tab:w}
\end{table}

Why is the $U_{Z,0}$ of ice II lower than that of ice Ih? $U_{Z,0}$
is proportional to the average of the vibrational frequencies, $\overline{\omega}_{k}$,
of the ice cell {[}see Eq. (\ref{eq:Uz}){]}. In Tab. \ref{tab:w},
the average of translational, librational, bending and stretching
modes is presented for the equilibrium cells of the ice phases. We
observe that the largest difference in $\overline{\omega}_{k}$ between
ice Ih and II is due to the librational modes, that are about 10\%
lower in ice II. The stretching modes show a competing behavior in
the sense that they are larger for ice II. But the overall effect
of all modes is the reduction of the zero-point energy of ice II in
comparison to ice Ih by about 0.67 kJ/mol (1\%). We can anticipate
that for rigid water models this competing mechanism between librational
and stretching modes is absent. Thus, the stabilization of ice II
in the quantum phase diagram of rigid models should be even larger
than in the flexible one.

\subsubsection{Coexistence II-III}

\begin{figure}
\vspace{-1.8cm}
\includegraphics[width= 9cm]{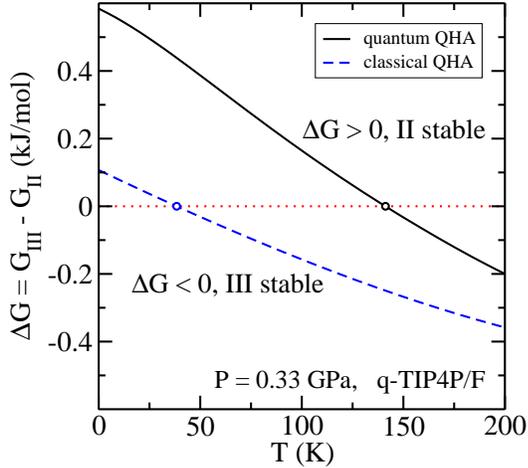}
\vspace{-0.9cm}
\caption{Gibbs free energy difference between ice III and II as a
function
of $T$ at $P=0.33$ GPa. Coexistence conditions are labeled by open
symbols ($\Delta G=0$). Both classical and quantum limits were derived
with the q-TIP4P/F model for the H-isomer of ice III labeled with
an arrow in Fig. \ref{fig:u0_v}.}
\label{fig:gII_III}
\end{figure}

The Gibbs free energy difference between ice III and II is plotted
as a function of the temperature in Fig. \ref{fig:gII_III}. Both
classical and quantum QHA limits are displayed at a constant pressure
of $P=0.33$ GPa. The coexistence temperature in the quantum limit
is shifted by about 100 K toward higher temperatures, i.e., quantum
effects play an important role in the stabilization of ice II. The
zero-point energies in Tab. \ref{tab:ices_ref_fle} show that $U_{Z}$
of ice II is significantly lower (about 0.6 kJ/mol) than that of ice
III. The slope of the $\Delta G$ curves in Fig. \ref{fig:gII_III}
is always negative 
\begin{equation}
\frac{\partial\left(\Delta G\right)}{\partial T}=-\Delta S=S_{II}-S_{III}<0\;,
\end{equation}
which is consistent with the larger entropy of ice III due to its
H-disorder. At a given temperature the slope of the quantum $\Delta G$
curve is larger (in absolute value) than in the classical result.
This implies that the excess of entropy of ice III, with respect to
ice II, is larger in the quantum limit, i.e., the quantum vibrational
entropy contributes to stabilize ice III. Nevertheless, the overall
quantum effect in $\Delta G$ implies a strong stabilization of ice
II with respect to ice III, as a consequence of its lower zero-point
energy.

\section{Rigid TIP4P/2005 model\label{sec:Rigid-TIP4P/2005-model}}

The QH phase diagram of ice Ih, II, and III with the rigid TIP4P/2005
potential has been calculated using the same H-isomers as those employed
for the q-TIP4P/F study in Fig. \ref{fig:Ae}. Differences in the
results should be attributed to the potential models (flexible versus
rigid) and not to effects related to the selected H-isomers. 

\begin{figure}
\vspace{-1.8cm}
\includegraphics[width= 9cm]{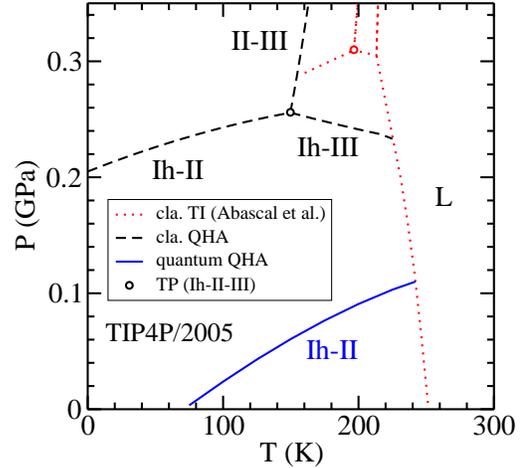}
\vspace{-0.9cm}
\caption{Comparison of the QH phase diagram of ice Ih, II, and III in
the classical
and quantum limits. Dotted lines correspond to the classical TI
simulations
of Ref. \onlinecite{abascal05} that include the liquid (L) phase.
These results were derived with the TIP4P/2005 model. Ice III is modeled
with the same H-isomer as that employed in Fig. \ref{fig:Ae} for
the q-TIP4P/F model. Circles show the position of the triple point
for ice Ih-II-III.}
\label{fig:R}
\end{figure}

The classical and quantum QH phase diagrams are presented in Fig.
\ref{fig:R} as dashed and continuous curves, respectively. The phase
diagram reported for the TIP4P/2005 model by classical TI simulations
is shown by dotted lines. In the last case the coexistence with the
liquid phase is also given. There are several aspects to be commented.
First is the comparison between the classical QHA and the TI results.
The coexistence lines between the ice phases are nearly parallel in
both calculations. We recall that the slope of the coexistence lines
is determined by the Clausius-Clapeyron relation
\begin{equation}
\frac{dP_{coe}}{dT_{coe}}=\frac{\Delta H}{T_{coe}\Delta V}=\frac{\Delta S}{\Delta V}\;,
\end{equation}
where $\Delta H$ is the enthalpy difference (latent heat) between
the two phases at equilibrium. It has been shown earlier that the
QHA provides a realistic approximation for the enthalpy and volume
of ice Ih, II, and III in a broad ($T,P$) interval.\cite{ramirez12}
Therefore, the QHA shows a reasonable overall agreement to the classical
TI results for the slopes of the coexistence lines. The largest deviation
between the QH and TI slopes is found for the Ih-II transition at
temperatures close to the triple point, where the slope of the QHA
is lower. The larger stability region of ice III in the QHA is likely
due to the finite size effect related to the different H-ordering
of the studied H-isomers, although the approximate treatment of anharmonic
effects by the QHA may be also the origin of this behavior.

The consideration of quantum vibrational effects changes dramatically
the QH phase diagram. The main quantum effect is the stabilization
of ice II. It has two important consequences. The first is that ice
II becomes the stable phase at low temperatures even at $P=0$. The
second is that ice III disappears as stable phase in the displayed
region of the phase diagram. Both facts imply that the quantum QH
phase diagram becomes in strong disagreement to experimental facts,
as opposed to the classical one. 

In the case of the TIP4P/2005 model there are not quantum TI results
available for comparison. Nevertheless the QH prediction that ice
II becomes the stable low temperature phase agrees with the quantum
path integral simulations of ice Ih and II in Ref. \onlinecite{mcbride09}.
The explanation given for this behavior was that the TIP4P/2005 model
is parameterized to be used in a classical limit. The combination
with quantum simulations implies some kind of overcounting of quantum
effects that leads to unphysical results. 

It is interesting to analyze the physical reason for the larger stabilization
of ice II in the quantum phase diagram of Fig. \ref{fig:R} (rigid
model) in comparison to Fig. \ref{fig:Ae} (flexible model). At $T=0$
and $P=0$ the shift in the coexistence pressure of ice Ih-II in the
classical and quantum limits is proportional to the difference in
the zero-point energy, $\Delta U_{Z,0}$, between both phases. For
TIP4P/2005 we get (see Tab. \ref{tab:ice_ref_2005} )
\begin{equation}
\Delta U_{Z.0}=U_{Z,0,Ih}-U_{Z,0,II}=1.05\;\mathrm{kJ/mol}\;,
\end{equation}
i.e., the zero-point energy of ice II is more than 1 kJ/mol lower
than that of ice Ih. This stabilization is about 50\% larger than
that of the flexible model {[}see Eq. (\ref{eq:d_uz}){]}. This fact
is a consequence of the absence in a rigid model of competing contributions
of librational and stretching modes to $\Delta U_{Z,0}$, as discussed
in Subsec. \ref{sub:Coexistence-Ih-II}. In conclusion, the stabilization
of ice II by its lower zero-point energy is larger for the rigid model
than for the flexible one. As a consequence ice II becomes the stable
phase at low temperature and ice III is unstable in the quantum limit
of the TIP4P/2005 model (see Fig. \ref{fig:R}).

A last remark on the comparison of the QH phase diagrams of the rigid
and flexible models. The most realistic results are obtained in the
quantum limit for the flexible model (see Fig. \ref{fig:Ae}), but
in the classical limit for the rigid model (see Fig. \ref{fig:R}).
The models were parameterized to be used in either quantum (the flexible
one)\cite{habershon09} or classical simulations (the rigid one).\cite{abascal05}
Thus, the best result correlates in each case with the conditions
where the model was parameterized.

\section{Rigid TIP4PQ/2005 model\label{sec:Rigid-TIP4PQ/2005-model}}

\begin{figure}
\vspace{-0.9cm}
\includegraphics[width= 9cm]{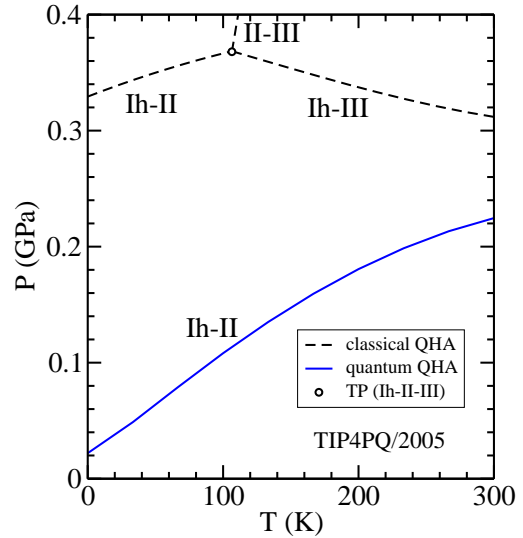}
\vspace{-0.9cm}
\caption{Comparison of the QHA phase diagram of ice Ih, II, and III in
the classical and quantum limits. Results derived with the TIP4PQ/2005
model. Ice III is modeled with the same H-isomer as that employed
in Fig. \ref{fig:R} for the TIP4P/2005 model. A circle shows the
position of the triple point for ice Ih-II-III in the classical limit.}
\label{fig:q1}
\end{figure}

The rigid TIP4PQ/2005 model differs from TIP4P/2005 by an increase
of about 4\% in the point charges, i.e., the parameter $q_{H}$ is
the only one that changes from 0.5564 $e$ to 0.5764 $e$.\cite{mcbride09}
The QH phase diagram of ice Ih, II, and III for the rigid TIP4PQ/2005
potential has been calculated with the same H-isomers as those used
in the studies shown in Fig. \ref{fig:Ae} (q-TIP4P/F) and Fig. \ref{fig:R}
(TIP4P/2005). The results for TIP4PQ/2005 are displayed in Fig. \ref{fig:q1}. 

We observe that the QH diagrams in Figs. \ref{fig:R} and \ref{fig:q1}
are qualitatively similar. In the classical case, there appears a
triple point for the ices Ih-II-III. However, in the quantum limit,
ice III is unstable and the phase diagram displays only the ice Ih-II
transition. Another similarity of both phase diagrams is that the
coexistence lines in Figs. \ref{fig:R} and \ref{fig:q1} are nearly
parallel. 

The change in the point charges modifies the static energy, $U_{S,ref},$
of the ice phases (see Tabs. \ref{tab:ice_ref_2005} and \ref{tab:ice_ref_Q2005}).
This translates into a shift of the coexistence lines calculated with
both rigid models. For example, in the quantum limit the transition
Ih-II appears at higher pressures for TIP4PQ/2005, so that ice Ih
becomes the stable low temperature phase (see Fig. \ref{fig:q1}).
This result is in agreement with the path integral simulations of
the TIP4PQ/2005 model in Ref. \onlinecite{mcbride09}. 

\begin{figure}
\vspace{-1.0cm}
\includegraphics[width= 9cm]{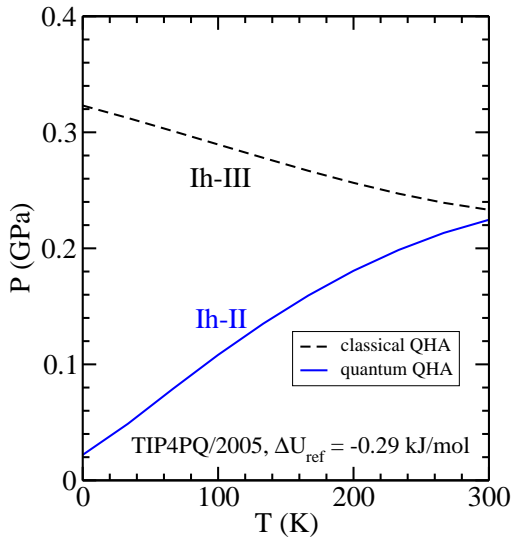}
\vspace{-0.9cm}
\caption{Comparison of the QH phase diagram of ice Ih, II, and III in
the classical
and quantum limits. Results derived with the TIP4PQ/2005 model. The
results differ from those in Fig. \ref{fig:q1} by an additional
stabilization
of ice III by a constant energy shift given by $\Delta U_{ref}$. }
\label{fig:q3}
\end{figure}

\begin{figure}
\vspace{-1.0cm}
\includegraphics[width= 9cm]{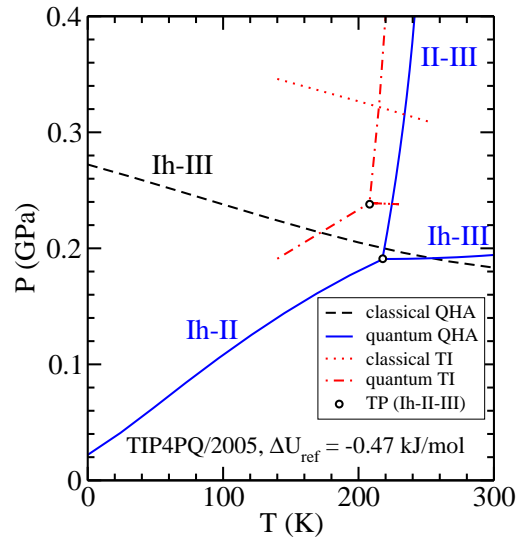}
\vspace{-0.9cm}
\caption{Comparison of the QH phase diagram of ice Ih, II, and III in
the classical
and quantum limits. Results derived with the TIP4PQ/2005 model. The
results differ from those in Fig. \ref{fig:q1} by an additional
stabilization
of ice III by a constant energy shift given by $\Delta U_{ref}$.
Dotted (dashed-dotted) lines correspond to the classical (quantum)
TI simulations of Ref. \onlinecite{mcbride12}. Circles show the position
of the triple point for ice Ih-II-III.}
\label{fig:q2}
\end{figure}

Classical and quantum phase diagrams of water, derived by TI simulations
with the TIP4PQ/2005 model, have been reported recently.\cite{mcbride12}
The results differ markedly from our QHA. We believe that the main
reason of discrepancy is the finite size effect in the value of the
static energy, $U_{S,ref}$, of ice III. As a check of this hypothesis,
we have repeated our QHA calculation of the TIP4PQ/2005 phase diagram
by shifting the internal energy of ice III by a constant amount. All
other model parameters remain unchanged (i.e., static energies of
ice Ih and II, and the vibrational properties of the three ice phases).
We have analyzed the effect of setting the static energy of ice III
as 
\begin{equation}
U_{S,new}(V)=U_{S}(V)+\Delta U_{ref}\;,
\end{equation}
where $\Delta U_{ref}$ is a constant energy shift. The QH phase diagram
for $\Delta U_{ref}=0$ was displayed in Fig. \ref{fig:q1}. The phase
diagrams derived for $\Delta U_{ref}=-0.29$ kJ/mol and $\Delta U_{ref}=-0.47$
kJ/mol are displayed in Figs. \ref{fig:q3} and \ref{fig:q2}, respectively.
The differences between these phase diagrams are caused by the artificial
stabilization of ice III by the constant energy $\Delta U_{ref}$.
Note that the selected energy shifts are in the order of the dispersion
of $U_{S,ref}$ represented in the abscissa of Fig. \ref{fig:u0_v}
for the q-TIP4P/F model.

In the quantum limit, the phase diagram calculated with $\Delta U_{ref}=-0.29$
kJ/mol (Fig. \ref{fig:q3}), is identical to that shown for $\Delta U_{ref}=0$
(Fig. \ref{fig:q1}). The coexistence Ih-II is the only transition.
The additional stabilization of ice III is not large enough to make
it stable in the quantum limit. However, it does change drastically
the classical limit of the phase diagram. Ice III becomes more stable
than ice II in the whole region, and the classical phase diagram loses
its triple point and now displays only the ice Ih-III transition.
Note that for $\Delta U_{ref}=-0.29$ kJ/mol, the triple point Ih-II-III
is missing in both classical and quantum limits.

Imposing a larger stabilization to ice III ($\Delta U_{ref}=-0.47$
kJ/mol), we obtain the QHA phase diagram shown in Fig. \ref{fig:q2}.
In this plot we have represented also the classical and quantum phase
diagrams for ice Ih, II, and III calculated by TI simulations in Ref.
\onlinecite{mcbride12}. We observe that this additional stabilization
brings the QHA in reasonable agreement to the TI results reported
for this model. Now a triple point Ih-II-III is observed in the quantum
limit, while the transition Ih-III is the only line in the classical
case. Furthermore, we note that the slopes of the coexistence lines
predicted by the QHA are in reasonable agreement to the TI results.

The sequence of phase diagrams shown in Figs. \ref{fig:q1}, \ref{fig:q3},
and \ref{fig:q2} provides a vivid illustration of the dramatic changes
in the ice phase diagram as a function of the stability of the employed
H-isomer of ice III.

\section{Average over proton disorder \label{sec:Average-over-proton}}

In this Section we comment on the necessity of performing some form
of proton disorder average, at least for ice III, prior to the calculation
of the phase diagram. Assuming that the number of H-isomers for an
ice cell is $M$, the canonical partition function of the ice phase
can be expressed as
\begin{equation}
e^{-\beta F}=\sum_{i=1}^{M}e^{-\beta(U_{S,i}+F_{v,i})}\;,\label{eq:average}
\end{equation}
where $U_{S,i}$ and $F_{v,i}$ are the static energy and the vibrational
free energy of \textit{i'}th H-isomer. Note that many of the $M$
H-isomers might be energetically degenerate as a consequence of the
lattice symmetry. Eq. (\ref{eq:average}) is the formally correct
way to average over the proton disorder of the ice phase. It has been
applied to study order-disorder transitions of ice phases using small
units cells.\cite{singer12} Obviously for large unit cells the total
number $M$ of H-isomers grows in such a way that the application
of Eq. (\ref{eq:average}) becomes an impossible task. 

The alternative for large unit cells is the use of Eq. (\ref{eq:f_v_t}),
which is applied to a \textit{single} H-isomer selected randomly from
the set of $M$ available ones.\cite{sanz04,abascal05,mcbride12,habershon11}
Note that in this equation the average over proton disorder is introduced
\textit{ad hoc} by the term with the proton disorder entropy $S_{H}$.
An implicit assumption in Eq. (\ref{eq:f_v_t}) is that the cell is
so large that the static energy, $U_{S},$ and the vibrational free
energy, $F_{v}$, of the single H-isomer do not require any further
average over the proton disorder. This assumption is correct in the
thermodynamic limit as the relative fluctuation of thermodynamic quantities
is expected to decrease as $1/\sqrt{N}$. 

However, we have shown that, for typical cell sizes used in simulations,
the fluctuation of $U_{S}$ for ice III is far from its ideal thermodynamic
limit. Therefore some form of proton disorder averaging of $U_{S}$
is necessary. A computational feasible proposal is to average only
the term that shows the largest fluctuation as a function of the H-disorder,
which is the potential energy of the reference cell $U_{S,ref}$ associated
to the chosen H-isomer. Thus, a simple proposal to average over proton
disorder is to modify the static energy in Eq. (\ref{eq:f_v_t}) by
\begin{equation}
U_{S,ave}(V)=U_{S}(V)+\Delta U_{ave}\;.
\end{equation}
$\Delta U_{ave}$ here is a constant energy shift that modifies the
stability of the single selected H-isomer of ice III by an amount
determined by the average $\overline{U}_{S,ref}$ calculated over
a random set of H-isomers, i.e., 
\begin{equation}
\Delta U_{ave}=\overline{U}_{S,ref}-U_{S,ref}\;.
\end{equation}
Note that the proposed disorder averaging can be applied to the calculation
of phase diagrams either by the QHA or by TI simulations.

The average of $U_{S,ref}$ over the set of six random H-isomers of
ice III with full H-disorder studied in this work gives $\overline{U}_{S,ref}=-60.96\pm0.04$
kJ/mol for the q-TIP4P/F model. We estimate that the error of the
mean value, $\overline{U}_{S,ref}$, should be as low as 0.01 kJ/mol
for a reasonable convergence over the proton disorder. Then the size
of our random sampling should be increased by a factor of 16 to reduce
our estimated error to this limit, i.e., for an ice III cell with
324 molecules one should increase the sampling of $U_{S,ref}$ to
about 100 H-isomers. By using larger units cells, e.g., a $4\times4\times4$
supercell with 768 water molecules, the number of required H-isomers
to obtained a converged value of $\overline{U}_{S,ref}$ should be
lower than this. However, in terms of computational efficiency, the
lower number of H-isomers may be overcompensated by the higher computational
cost in the minimization of the cell energies.

In the case of the partially disordered ice III we get $\overline{U}_{S,ref}=-60.73\pm0.01$
kJ/mol for our set of six random H-isomers with the q-TIP4P/F model.
This value has achieved already the desired convergence. A last comment
is that the proton disorder entropy $S_{H}$ for partially disordered
phases is lower than the Pauling estimate in Eq. (\ref{eq:entropy}).
The estimation of $S_{H}$ for the fractional H-occupancies experimentally
determined for ice III amounts to about 90\% of the Pauling result.
This entropy lowering has a significant influence in the phase diagram.\cite{macdowell04}
Numerical methods for the estimation of the proton disorder entropy
of partially H-disordered phases can be found in Refs. \onlinecite{macdowell04,berg07b}.

\section{Conclusions\label{sec:Conclusions}}

In this work we have presented a detailed study of the phase diagram
of ice Ih, II, and III calculated with the QHA and TIP4P-like models.
Several advantages of the QHA are worth to be mentioned: its computational
cost is low, it is free from the statistical errors inherent to any
numerical simulation, it can be applied to both classical and quantum
limits, and the most accurate results are expected in the low temperature
limit, where anharmonicities are lower. These advantages make the
QHA an appropriate option to study finite size effects that can become
prohibitively expensive in numerical (Monte Carlo or molecular dynamics)
simulations.

The effect of proton disorder in the phase diagram of TIP4P-like models
has been a focus of our study. We have found that for the typical
cell sizes employed in computer simulations, the finite size effect
of H-disorder in ice Ih is very small. However, this effect is very
large for ice III, so that the transitions II-III and Ih-III are strongly
affected by it. The physical reason for this behavior is that the
static energy of ice III may change by an amount of several tenths
of kJ/mol depending on the considered H-configuration. Thus a randomly
selected ice III structure makes the calculated phase diagram affected
by an uncontrolled factor that can be highly significant for the final
result. Crystallographic data of ice Ih and III reveal significant
differences as a consequence of the larger structural complexity of
ice III. As example, ice Ih is a network of hexagon rings of O-atoms,
while in ice III there appear five-, seven, and eight-members rings.
The tetragonal O-O-O angles deform from the ideal value of about 109$^{\circ}$
in ice Ih into angles between 80$^{\circ}$ and 140$^{\circ}$ in
ice III.\cite{lobban00} Thus the large finite size effect related
to the H-disorder in ice III, in comparison to ice Ih, seems to correlate
with its increased structural complexity. We have discussed the necessity
of performing a disorder average of the lattice energy in order to
reduce this effect.

Another aspect related to the H-disorder in ice III is that TIP4P-like
models predict that full H-disorder is more stable than partial H-disorder.
However, this result is against the data derived from diffraction
experiments of ice III.\cite{lobban00} A phase diagram derived with
ice III having partial H-disorder may be significantly different from
that derived with full H-disorder, as a consequence of the differences
in their static energies. 

These findings allows us to rationalize previously contradictory results
of phase diagrams calculated with TIP4P-like potentials. A significant
example was the reported instability of ice II in the classical limit
of the flexible q-TIP4P/F model.\cite{habershon11} This fact contrasts
with the stability of ice II reported with rigid TIP4P and TIP4P/2005
models.\cite{sanz04,abascal05} Our QH results show that the stability
of ice II strongly depends upon the H-configuration of the chosen
ice III isomer. The fact that ice III has been described either with
partial\cite{sanz04} or full H-disorder\cite{habershon11,mcbride12}
makes difficult the comparison of phase diagrams with different models.
It is not easy to discriminate from the reported simulations the effects
due to the variable stability of the ice III isomers from those caused
by the different water models.

The QH phase diagram of ices Ih-II-III for the q-TIP4P/F, TIP4P/2005
and TIP4PQ/2005 effective potentials has been able to reproduce qualitatively
most features that had been previously studied by TI in both classical
and quantum simulations. Our results are obtained using the same H-isomers
for the three TIP4P-like models, allowing for an easier interpretation
of the differences encountered in the phase diagrams. We have found
that for the flexible model (q-TIP4P/F) the triple point Ih-II-III
is shifted in the quantum limit by 100 K and -0.06 GPa with respect
to the classical one. This effect is very large, specially in the
temperature. Its physical origin is related to the lower zero-point
energy of ice II, when compared to that of ice Ih and III. This fact
translates in an increased stability of ice II when vibrational quantum
effects are considered. The average frequency of molecular librations
in the H-bond network are nearly 10\% smaller in ice II than in the
other ice phases. This causes a significant reduction of the zero-point
energy. Interestingly, the intramolecular stretching modes of ice
II are predicted at higher frequencies than those in ice Ih and III.
This anticorrelation between libration and stretching modes has been
often stressed in the literature, i.e., any factor that shifts librational
frequencies in one direction acts also modifying the stretching frequencies
in the opposite one.\cite{pamuk12} We find as net effect that librational
modes dominate over stretching ones, leading to the stabilization
of ice II by its lower zero-point energy.

The main difference between the QH phase diagrams obtained by the
flexible and rigid models is associated to the absence of anticorrelation
between H-bond librations and O-H stretchings in the rigid models.
Obviously intramolecular bonds are frozen for rigid water. As a consequence,
the stabilization of ice II by its zero-point energy is larger for
the rigid models (TIP4P/2005, TIP4PQ/2005) than for the flexible one
(q-TIP4P/F). It may be even so large that ice II becomes the stable
phase at low temperatures. This result has been reported by quantum
path integral simulations with the TIP4P/2005 potential in Ref. \onlinecite{abascal05}
and is also reproduced by our QHA. Another effect related to the large
stabilization of ice II is that this phase may occupy the stability
region of ice III in the quantum limit of the rigid models. Then the
triple point Ih-II-III does not appear in the quantum phase diagram.
This unphysical behavior, as displayed in the quantum phase diagrams
of Figs. \ref{fig:R} and \ref{fig:q1}, can be avoided if the H-isomer
of ice III is particularly stable, as was shown in the phase diagram
of Fig. \ref{fig:q2}.

The present work can be extended along several directions. An obvious
one is the analysis of the phase diagram of other ice phases with
the flexible q-TIP4P/F model as well as the study of finite size effects
in the proton disorder of other phases, as ice V, VI, and VII. A second
aim is the use of DFT to avoid the limitations of the empirical potentials,
in particular with respect to the energetics associated to proton
rearrangements in ice phases. 

\acknowledgments

This work was supported by Ministerio de Ciencia e Innovación (Spain)
through Grant No. FIS2009-12721-C04-04, and by Comunidad Autónoma
de Madrid through project MODELICO-CM/S2009ESP-1691. We thank M.-V.
Fernández-Serra for insightful discussions. 

\bibliographystyle{apsrev}

\begin{thebibliography}{38}
\expandafter\ifx\csname natexlab\endcsname\relax\def\natexlab#1{#1}\fi
\expandafter\ifx\csname bibnamefont\endcsname\relax
  \def\bibnamefont#1{#1}\fi
\expandafter\ifx\csname bibfnamefont\endcsname\relax
  \def\bibfnamefont#1{#1}\fi
\expandafter\ifx\csname citenamefont\endcsname\relax
  \def\citenamefont#1{#1}\fi
\expandafter\ifx\csname url\endcsname\relax
  \def\url#1{\texttt{#1}}\fi
\expandafter\ifx\csname urlprefix\endcsname\relax\def\urlprefix{URL }\fi
\providecommand{\bibinfo}[2]{#2}
\providecommand{\eprint}[2][]{\url{#2}}

\bibitem[{\citenamefont{Dunaeva et~al.}(2010)\citenamefont{Dunaeva, Antsyshkin,
  and Kuskov}}]{dunaeva10}
\bibinfo{author}{\bibfnamefont{A.}~\bibnamefont{Dunaeva}},
  \bibinfo{author}{\bibfnamefont{D.}~\bibnamefont{Antsyshkin}},
  \bibnamefont{and} \bibinfo{author}{\bibfnamefont{O.}~\bibnamefont{Kuskov}},
  \bibinfo{journal}{Solar System Research} \textbf{\bibinfo{volume}{44}},
  \bibinfo{pages}{202} (\bibinfo{year}{2010}).

\bibitem[{\citenamefont{Bernal and Fowler}(1933)}]{bernal33}
\bibinfo{author}{\bibfnamefont{J.~D.} \bibnamefont{Bernal}} \bibnamefont{and}
  \bibinfo{author}{\bibfnamefont{R.~H.} \bibnamefont{Fowler}},
  \bibinfo{journal}{J. Chem. Phys.} \textbf{\bibinfo{volume}{1}},
  \bibinfo{pages}{515} (\bibinfo{year}{1933}).

\bibitem[{\citenamefont{Singer and Knight}(2011)}]{singer12}
\bibinfo{author}{\bibfnamefont{S.~J.} \bibnamefont{Singer}} \bibnamefont{and}
  \bibinfo{author}{\bibfnamefont{C.}~\bibnamefont{Knight}},
  \bibinfo{journal}{Adv. Chem. Phys.} \textbf{\bibinfo{volume}{147}},
  \bibinfo{pages}{1} (\bibinfo{year}{2011}).

\bibitem[{\citenamefont{Sanz et~al.}(2004)\citenamefont{Sanz, Vega, Abascal,
  and MacDowell}}]{sanz04}
\bibinfo{author}{\bibfnamefont{E.}~\bibnamefont{Sanz}},
  \bibinfo{author}{\bibfnamefont{C.}~\bibnamefont{Vega}},
  \bibinfo{author}{\bibfnamefont{J.~L.~F.} \bibnamefont{Abascal}},
  \bibnamefont{and} \bibinfo{author}{\bibfnamefont{L.~G.}
  \bibnamefont{MacDowell}}, \bibinfo{journal}{Phys. Rev. Lett.}
  \textbf{\bibinfo{volume}{92}}, \bibinfo{pages}{255701}
  (\bibinfo{year}{2004}).

\bibitem[{\citenamefont{Abascal and Vega}(2005)}]{abascal05}
\bibinfo{author}{\bibfnamefont{J.~L.~F.} \bibnamefont{Abascal}}
  \bibnamefont{and} \bibinfo{author}{\bibfnamefont{C.}~\bibnamefont{Vega}},
  \bibinfo{journal}{J. Chem. Phys.} \textbf{\bibinfo{volume}{123}},
  \bibinfo{eid}{234505} (\bibinfo{year}{2005}).

\bibitem[{\citenamefont{McBride et~al.}(2012)\citenamefont{McBride, Noya,
  Aragones, Conde, and Vega}}]{mcbride12}
\bibinfo{author}{\bibfnamefont{C.}~\bibnamefont{McBride}},
  \bibinfo{author}{\bibfnamefont{E.~G.} \bibnamefont{Noya}},
  \bibinfo{author}{\bibfnamefont{J.~L.} \bibnamefont{Aragones}},
  \bibinfo{author}{\bibfnamefont{M.}~\bibnamefont{Conde}}, \bibnamefont{and}
  \bibinfo{author}{\bibfnamefont{C.}~\bibnamefont{Vega}},
  \bibinfo{journal}{Phys. Chem. Chem. Phys.} \textbf{\bibinfo{volume}{14}},
  \bibinfo{pages}{10140} (\bibinfo{year}{2012}).

\bibitem[{\citenamefont{Habershon and
  Manolopoulos}(2011{\natexlab{a}})}]{habershon11}
\bibinfo{author}{\bibfnamefont{S.}~\bibnamefont{Habershon}} \bibnamefont{and}
  \bibinfo{author}{\bibfnamefont{D.~E.} \bibnamefont{Manolopoulos}},
  \bibinfo{journal}{Phys. Chem. Chem. Phys.} \textbf{\bibinfo{volume}{13}},
  \bibinfo{pages}{19714} (\bibinfo{year}{2011}{\natexlab{a}}).

\bibitem[{\citenamefont{Jorgensen et~al.}(1983)\citenamefont{Jorgensen,
  Chandrasekhar, Madura, Impey, and Klein}}]{jorgensen83}
\bibinfo{author}{\bibfnamefont{W.~L.} \bibnamefont{Jorgensen}},
  \bibinfo{author}{\bibfnamefont{J.}~\bibnamefont{Chandrasekhar}},
  \bibinfo{author}{\bibfnamefont{J.~D.} \bibnamefont{Madura}},
  \bibinfo{author}{\bibfnamefont{R.~W.} \bibnamefont{Impey}}, \bibnamefont{and}
  \bibinfo{author}{\bibfnamefont{M.~L.} \bibnamefont{Klein}},
  \bibinfo{journal}{J. Chem. Phys.} \textbf{\bibinfo{volume}{79}},
  \bibinfo{pages}{926} (\bibinfo{year}{1983}).

\bibitem[{\citenamefont{McBride et~al.}(2009)\citenamefont{McBride, Vega, Noya,
  Ram\'{\i}rez, and Ses\'{e}}}]{mcbride09}
\bibinfo{author}{\bibfnamefont{C.}~\bibnamefont{McBride}},
  \bibinfo{author}{\bibfnamefont{C.}~\bibnamefont{Vega}},
  \bibinfo{author}{\bibfnamefont{E.~G.} \bibnamefont{Noya}},
  \bibinfo{author}{\bibfnamefont{R.}~\bibnamefont{Ram\'{\i}rez}},
  \bibnamefont{and} \bibinfo{author}{\bibfnamefont{L.~M.}
  \bibnamefont{Ses\'{e}}}, \bibinfo{journal}{J. Chem. Phys.}
  \textbf{\bibinfo{volume}{131}}, \bibinfo{eid}{024506} (\bibinfo{year}{2009}).

\bibitem[{\citenamefont{Habershon et~al.}(2009)\citenamefont{Habershon,
  Markland, and Manolopoulos}}]{habershon09}
\bibinfo{author}{\bibfnamefont{S.}~\bibnamefont{Habershon}},
  \bibinfo{author}{\bibfnamefont{T.~E.} \bibnamefont{Markland}},
  \bibnamefont{and} \bibinfo{author}{\bibfnamefont{D.~E.}
  \bibnamefont{Manolopoulos}}, \bibinfo{journal}{J. Chem. Phys.}
  \textbf{\bibinfo{volume}{131}}, \bibinfo{eid}{024501} (\bibinfo{year}{2009}).

\bibitem[{\citenamefont{Vega et~al.}(2008)\citenamefont{Vega, Sanz, Abascal,
  and Noya}}]{vega09}
\bibinfo{author}{\bibfnamefont{C.}~\bibnamefont{Vega}},
  \bibinfo{author}{\bibfnamefont{E.}~\bibnamefont{Sanz}},
  \bibinfo{author}{\bibfnamefont{J.~L.~F.} \bibnamefont{Abascal}},
  \bibnamefont{and} \bibinfo{author}{\bibfnamefont{E.~G.} \bibnamefont{Noya}},
  \bibinfo{journal}{J. Phys.: Condens. Matter} \textbf{\bibinfo{volume}{20}},
  \bibinfo{pages}{153101} (\bibinfo{year}{2008}).

\bibitem[{\citenamefont{Buch et~al.}(1998)\citenamefont{Buch, Sandler, and
  Sadlej}}]{buch98}
\bibinfo{author}{\bibfnamefont{V.}~\bibnamefont{Buch}},
  \bibinfo{author}{\bibfnamefont{P.}~\bibnamefont{Sandler}}, \bibnamefont{and}
  \bibinfo{author}{\bibfnamefont{J.}~\bibnamefont{Sadlej}},
  \bibinfo{journal}{J. Chem. Phys} \textbf{\bibinfo{volume}{102}},
  \bibinfo{pages}{8641} (\bibinfo{year}{1998}).

\bibitem[{\citenamefont{Dion et~al.}(2004)\citenamefont{Dion, Rydberg,
  Schr\"oder, Langreth, and Lundqvist}}]{dion04}
\bibinfo{author}{\bibfnamefont{M.}~\bibnamefont{Dion}},
  \bibinfo{author}{\bibfnamefont{H.}~\bibnamefont{Rydberg}},
  \bibinfo{author}{\bibfnamefont{E.}~\bibnamefont{Schr\"oder}},
  \bibinfo{author}{\bibfnamefont{D.~C.} \bibnamefont{Langreth}},
  \bibnamefont{and} \bibinfo{author}{\bibfnamefont{B.~I.}
  \bibnamefont{Lundqvist}}, \bibinfo{journal}{Phys. Rev. Lett.}
  \textbf{\bibinfo{volume}{92}}, \bibinfo{pages}{246401}
  (\bibinfo{year}{2004}).

\bibitem[{\citenamefont{Wang et~al.}(2011)\citenamefont{Wang,
  Rom\'{a}n-P\'{e}rez, Soler, Artacho, and Fern\'{a}ndez-Serra}}]{wang11}
\bibinfo{author}{\bibfnamefont{J.}~\bibnamefont{Wang}},
  \bibinfo{author}{\bibfnamefont{G.}~\bibnamefont{Rom\'{a}n-P\'{e}rez}},
  \bibinfo{author}{\bibfnamefont{J.~M.} \bibnamefont{Soler}},
  \bibinfo{author}{\bibfnamefont{E.}~\bibnamefont{Artacho}}, \bibnamefont{and}
  \bibinfo{author}{\bibfnamefont{M.-V.} \bibnamefont{Fern\'{a}ndez-Serra}},
  \bibinfo{journal}{J. Chem. Phys.} \textbf{\bibinfo{volume}{134}},
  \bibinfo{pages}{024516} (\bibinfo{year}{2011}).

\bibitem[{\citenamefont{Santra et~al.}(2011)\citenamefont{Santra,
  Klime\ifmmode~\check{s}\else \v{s}\fi{}, Alf\`e, Tkatchenko, Slater,
  Michaelides, Car, and Scheffler}}]{santra11}
\bibinfo{author}{\bibfnamefont{B.}~\bibnamefont{Santra}},
  \bibinfo{author}{\bibfnamefont{J.}~\bibnamefont{Klime\ifmmode~\check{s}\else
  \v{s}\fi{}}}, \bibinfo{author}{\bibfnamefont{D.}~\bibnamefont{Alf\`e}},
  \bibinfo{author}{\bibfnamefont{A.}~\bibnamefont{Tkatchenko}},
  \bibinfo{author}{\bibfnamefont{B.}~\bibnamefont{Slater}},
  \bibinfo{author}{\bibfnamefont{A.}~\bibnamefont{Michaelides}},
  \bibinfo{author}{\bibfnamefont{R.}~\bibnamefont{Car}}, \bibnamefont{and}
  \bibinfo{author}{\bibfnamefont{M.}~\bibnamefont{Scheffler}},
  \bibinfo{journal}{Phys. Rev. Lett.} \textbf{\bibinfo{volume}{107}},
  \bibinfo{pages}{185701} (\bibinfo{year}{2011}).

\bibitem[{\citenamefont{Srivastava}(1990)}]{srivastava}
\bibinfo{author}{\bibfnamefont{G.~P.} \bibnamefont{Srivastava}},
  \emph{\bibinfo{title}{The Physics of Phonons}} (\bibinfo{publisher}{Adam
  Hilger}, \bibinfo{address}{Bristol}, \bibinfo{year}{1990}).

\bibitem[{\citenamefont{Pamuk et~al.}(2012)\citenamefont{Pamuk, Soler,
  Ram\'{i}rez, Herrero, Stephens, Allen, and Fern\'{a}ndez-Serra}}]{pamuk12}
\bibinfo{author}{\bibfnamefont{B.}~\bibnamefont{Pamuk}},
  \bibinfo{author}{\bibfnamefont{J.~M.} \bibnamefont{Soler}},
  \bibinfo{author}{\bibfnamefont{R.}~\bibnamefont{Ram\'{i}rez}},
  \bibinfo{author}{\bibfnamefont{C.~P.} \bibnamefont{Herrero}},
  \bibinfo{author}{\bibfnamefont{P.~W.} \bibnamefont{Stephens}},
  \bibinfo{author}{\bibfnamefont{P.~B.} \bibnamefont{Allen}}, \bibnamefont{and}
  \bibinfo{author}{\bibfnamefont{M.-V.} \bibnamefont{Fern\'{a}ndez-Serra}},
  \bibinfo{journal}{Phys. Rev. Lett.} \textbf{\bibinfo{volume}{108}},
  \bibinfo{pages}{193003} (\bibinfo{year}{2012}).

\bibitem[{\citenamefont{Tanaka}(2001)}]{tanaka01}
\bibinfo{author}{\bibfnamefont{H.}~\bibnamefont{Tanaka}}, \bibinfo{journal}{J.
  Mol. Liquids} \textbf{\bibinfo{volume}{90}}, \bibinfo{pages}{323 }
  (\bibinfo{year}{2001}).

\bibitem[{\citenamefont{Tse et~al.}(1999{\natexlab{a}})\citenamefont{Tse,
  Shpakov, and Belosludov}}]{tse99}
\bibinfo{author}{\bibfnamefont{J.~S.} \bibnamefont{Tse}},
  \bibinfo{author}{\bibfnamefont{V.~P.} \bibnamefont{Shpakov}},
  \bibnamefont{and} \bibinfo{author}{\bibfnamefont{V.~R.}
  \bibnamefont{Belosludov}}, \bibinfo{journal}{J. Chem. Phys.}
  \textbf{\bibinfo{volume}{111}}, \bibinfo{pages}{11111}
  (\bibinfo{year}{1999}{\natexlab{a}}).

\bibitem[{\citenamefont{Tse et~al.}(1999{\natexlab{b}})\citenamefont{Tse, Klug,
  Tulk, Swainson, Svensson, Loong, Shpakov, Belosludov, Belosludov, and
  Kawazoe}}]{tse99b}
\bibinfo{author}{\bibfnamefont{J.}~\bibnamefont{Tse}},
  \bibinfo{author}{\bibfnamefont{D.~D.} \bibnamefont{Klug}},
  \bibinfo{author}{\bibfnamefont{C.~A.} \bibnamefont{Tulk}},
  \bibinfo{author}{\bibfnamefont{I.~P.} \bibnamefont{Swainson}},
  \bibinfo{author}{\bibfnamefont{E.~C.} \bibnamefont{Svensson}},
  \bibinfo{author}{\bibfnamefont{C.-K.} \bibnamefont{Loong}},
  \bibinfo{author}{\bibfnamefont{V.~P.} \bibnamefont{Shpakov}},
  \bibinfo{author}{\bibfnamefont{V.~R.} \bibnamefont{Belosludov}},
  \bibinfo{author}{\bibfnamefont{R.~V.} \bibnamefont{Belosludov}},
  \bibnamefont{and} \bibinfo{author}{\bibfnamefont{Y.}~\bibnamefont{Kawazoe}},
  \bibinfo{journal}{Nature} \textbf{\bibinfo{volume}{400}},
  \bibinfo{pages}{647} (\bibinfo{year}{1999}{\natexlab{b}}).

\bibitem[{\citenamefont{Umemoto et~al.}(2010)\citenamefont{Umemoto,
  Wentzcovitch, de~Gironcoli, and Baroni}}]{umemoto10}
\bibinfo{author}{\bibfnamefont{K.}~\bibnamefont{Umemoto}},
  \bibinfo{author}{\bibfnamefont{R.~M.} \bibnamefont{Wentzcovitch}},
  \bibinfo{author}{\bibfnamefont{S.}~\bibnamefont{de~Gironcoli}},
  \bibnamefont{and} \bibinfo{author}{\bibfnamefont{S.}~\bibnamefont{Baroni}},
  \bibinfo{journal}{Chem. Phys. Lett.} \textbf{\bibinfo{volume}{499}},
  \bibinfo{pages}{236 } (\bibinfo{year}{2010}).

\bibitem[{\citenamefont{Ram\'{i}rez et~al.}(2012)\citenamefont{Ram\'{i}rez,
  Neuerburg, Fern\'{a}ndez-Serra, and Herrero}}]{ramirez12}
\bibinfo{author}{\bibfnamefont{R.}~\bibnamefont{Ram\'{i}rez}},
  \bibinfo{author}{\bibfnamefont{N.}~\bibnamefont{Neuerburg}},
  \bibinfo{author}{\bibfnamefont{M.-V.} \bibnamefont{Fern\'{a}ndez-Serra}},
  \bibnamefont{and} \bibinfo{author}{\bibfnamefont{C.~P.}
  \bibnamefont{Herrero}}, \bibinfo{journal}{J. Chem. Phys.}
  \textbf{\bibinfo{volume}{137}}, \bibinfo{pages}{044502}
  (\bibinfo{year}{2012}).

\bibitem[{\citenamefont{Koyama et~al.}(2004)\citenamefont{Koyama, Tanaka, Gao,
  and Zeng}}]{koyama04}
\bibinfo{author}{\bibfnamefont{Y.}~\bibnamefont{Koyama}},
  \bibinfo{author}{\bibfnamefont{H.}~\bibnamefont{Tanaka}},
  \bibinfo{author}{\bibfnamefont{G.}~\bibnamefont{Gao}}, \bibnamefont{and}
  \bibinfo{author}{\bibfnamefont{X.~C.} \bibnamefont{Zeng}},
  \bibinfo{journal}{J. Chem. Phys.} \textbf{\bibinfo{volume}{121}},
  \bibinfo{pages}{7926} (\bibinfo{year}{2004}).

\bibitem[{\citenamefont{Noya et~al.}(1997)\citenamefont{Noya, Herrero, and
  Ram\'{\i}rez}}]{noya97}
\bibinfo{author}{\bibfnamefont{J.~C.} \bibnamefont{Noya}},
  \bibinfo{author}{\bibfnamefont{C.~P.} \bibnamefont{Herrero}},
  \bibnamefont{and}
  \bibinfo{author}{\bibfnamefont{R.}~\bibnamefont{Ram\'{\i}rez}},
  \bibinfo{journal}{Phys. Rev. B} \textbf{\bibinfo{volume}{56}},
  \bibinfo{pages}{237} (\bibinfo{year}{1997}).

\bibitem[{\citenamefont{Herrero and Ram\'{\i}rez}(2000)}]{herrero00}
\bibinfo{author}{\bibfnamefont{C.~P.} \bibnamefont{Herrero}} \bibnamefont{and}
  \bibinfo{author}{\bibfnamefont{R.}~\bibnamefont{Ram\'{\i}rez}},
  \bibinfo{journal}{Phys. Rev. B} \textbf{\bibinfo{volume}{63}},
  \bibinfo{pages}{024103} (\bibinfo{year}{2000}).

\bibitem[{\citenamefont{Herrero and Ram\'{\i}rez}(2005)}]{herrero05}
\bibinfo{author}{\bibfnamefont{C.~P.} \bibnamefont{Herrero}} \bibnamefont{and}
  \bibinfo{author}{\bibfnamefont{R.}~\bibnamefont{Ram\'{\i}rez}},
  \bibinfo{journal}{Phys. Rev. B} \textbf{\bibinfo{volume}{71}},
  \bibinfo{pages}{174111} (\bibinfo{year}{2005}).

\bibitem[{\citenamefont{Pauling}(1935)}]{pauling35}
\bibinfo{author}{\bibfnamefont{L.}~\bibnamefont{Pauling}}, \bibinfo{journal}{J.
  Am. Chem. Soc.} \textbf{\bibinfo{volume}{57}}, \bibinfo{pages}{2680}
  (\bibinfo{year}{1935}).

\bibitem[{\citenamefont{Kresse et~al.}(1995)\citenamefont{Kresse,
  Furthm\"uller, and Hafner}}]{kresse95}
\bibinfo{author}{\bibfnamefont{G.}~\bibnamefont{Kresse}},
  \bibinfo{author}{\bibfnamefont{J.}~\bibnamefont{Furthm\"uller}},
  \bibnamefont{and} \bibinfo{author}{\bibfnamefont{J.}~\bibnamefont{Hafner}},
  \bibinfo{journal}{Europhys. Lett.} \textbf{\bibinfo{volume}{32}},
  \bibinfo{pages}{729} (\bibinfo{year}{1995}).

\bibitem[{\citenamefont{Alf\`e et~al.}(2001)\citenamefont{Alf\`e, Price, and
  Gillan}}]{alfe01}
\bibinfo{author}{\bibfnamefont{D.}~\bibnamefont{Alf\`e}},
  \bibinfo{author}{\bibfnamefont{G.~D.} \bibnamefont{Price}}, \bibnamefont{and}
  \bibinfo{author}{\bibfnamefont{M.~J.} \bibnamefont{Gillan}},
  \bibinfo{journal}{Phys. Rev. B} \textbf{\bibinfo{volume}{64}},
  \bibinfo{pages}{045123} (\bibinfo{year}{2001}).

\bibitem[{\citenamefont{Venkataraman and Sahni}(1970)}]{venkataraman70}
\bibinfo{author}{\bibfnamefont{G.}~\bibnamefont{Venkataraman}}
  \bibnamefont{and} \bibinfo{author}{\bibfnamefont{V.~C.} \bibnamefont{Sahni}},
  \bibinfo{journal}{Rev. Mod. Phys.} \textbf{\bibinfo{volume}{42}},
  \bibinfo{pages}{409} (\bibinfo{year}{1970}).

\bibitem[{\citenamefont{Johnson et~al.}(1993)\citenamefont{Johnson, Zollweg,
  and Gubbins}}]{johnson93}
\bibinfo{author}{\bibfnamefont{J.~K.} \bibnamefont{Johnson}},
  \bibinfo{author}{\bibfnamefont{J.~A.} \bibnamefont{Zollweg}},
  \bibnamefont{and} \bibinfo{author}{\bibfnamefont{K.~E.}
  \bibnamefont{Gubbins}}, \bibinfo{journal}{Mol. Phys.}
  \textbf{\bibinfo{volume}{78}}, \bibinfo{pages}{591} (\bibinfo{year}{1993}).

\bibitem[{\citenamefont{Hayward and Reimers}(1997)}]{hayward87}
\bibinfo{author}{\bibfnamefont{J.~A.} \bibnamefont{Hayward}} \bibnamefont{and}
  \bibinfo{author}{\bibfnamefont{J.~R.} \bibnamefont{Reimers}},
  \bibinfo{journal}{J. Chem. Phys} \textbf{\bibinfo{volume}{106}},
  \bibinfo{pages}{1518} (\bibinfo{year}{1997}).

\bibitem[{\citenamefont{Kamb et~al.}(1971)\citenamefont{Kamb, Hamilton,
  LaPlaca, and Prakash}}]{kamb71}
\bibinfo{author}{\bibfnamefont{B.}~\bibnamefont{Kamb}},
  \bibinfo{author}{\bibfnamefont{W.~C.} \bibnamefont{Hamilton}},
  \bibinfo{author}{\bibfnamefont{S.~J.} \bibnamefont{LaPlaca}},
  \bibnamefont{and} \bibinfo{author}{\bibfnamefont{A.}~\bibnamefont{Prakash}},
  \bibinfo{journal}{J. Chem. Phys} \textbf{\bibinfo{volume}{55}},
  \bibinfo{pages}{1934} (\bibinfo{year}{1971}).

\bibitem[{\citenamefont{Lobban et~al.}(2000)\citenamefont{Lobban, Finney, and
  Kuhs}}]{lobban00}
\bibinfo{author}{\bibfnamefont{C.}~\bibnamefont{Lobban}},
  \bibinfo{author}{\bibfnamefont{J.~L.} \bibnamefont{Finney}},
  \bibnamefont{and} \bibinfo{author}{\bibfnamefont{W.~F.} \bibnamefont{Kuhs}},
  \bibinfo{journal}{J. Chem. Phys} \textbf{\bibinfo{volume}{112}},
  \bibinfo{pages}{7169} (\bibinfo{year}{2000}).

\bibitem[{\citenamefont{MacDowell et~al.}(2004)\citenamefont{MacDowell, Sanz,
  Vega, and Abascal}}]{macdowell04}
\bibinfo{author}{\bibfnamefont{L.~G.} \bibnamefont{MacDowell}},
  \bibinfo{author}{\bibfnamefont{E.}~\bibnamefont{Sanz}},
  \bibinfo{author}{\bibfnamefont{C.}~\bibnamefont{Vega}}, \bibnamefont{and}
  \bibinfo{author}{\bibfnamefont{J.~L.~F.} \bibnamefont{Abascal}},
  \bibinfo{journal}{J. Chem. Phys.} \textbf{\bibinfo{volume}{121}},
  \bibinfo{pages}{10145} (\bibinfo{year}{2004}).

\bibitem[{\citenamefont{Ram\'{\i}rez and Herrero}(2010)}]{ramirez10}
\bibinfo{author}{\bibfnamefont{R.}~\bibnamefont{Ram\'{\i}rez}}
  \bibnamefont{and} \bibinfo{author}{\bibfnamefont{C.~P.}
  \bibnamefont{Herrero}}, \bibinfo{journal}{J. Chem. Phys.}
  \textbf{\bibinfo{volume}{133}}, \bibinfo{eid}{144511} (\bibinfo{year}{2010}).

\bibitem[{\citenamefont{Habershon and
  Manolopoulos}(2011{\natexlab{b}})}]{habershon11b}
\bibinfo{author}{\bibfnamefont{S.}~\bibnamefont{Habershon}} \bibnamefont{and}
  \bibinfo{author}{\bibfnamefont{D.~E.} \bibnamefont{Manolopoulos}},
  \bibinfo{journal}{J. Chem. Phys.} \textbf{\bibinfo{volume}{135}},
  \bibinfo{eid}{224111} (\bibinfo{year}{2011}{\natexlab{b}}).

\bibitem[{\citenamefont{Berg and Yang}(2007)}]{berg07b}
\bibinfo{author}{\bibfnamefont{B.~A.} \bibnamefont{Berg}} \bibnamefont{and}
  \bibinfo{author}{\bibfnamefont{W.}~\bibnamefont{Yang}}, \bibinfo{journal}{J.
  Chem. Phys.} \textbf{\bibinfo{volume}{127}}, \bibinfo{eid}{224502}
  (\bibinfo{year}{2007}).

\end{thebibliography}

\end{document}